\documentclass[a4paper,11pt]{article}
\pdfoutput=1 

\usepackage{jheppub} 

\usepackage[T1]{fontenc} 
\usepackage{booktabs}
\usepackage{ifthen} 
\usepackage{tikz}
\usetikzlibrary{arrows}

\newboolean{pdflatex}
\setboolean{pdflatex}{true} 

\newboolean{uprightparticles}
\setboolean{uprightparticles}{false} 

\title{\boldmath Towards the precision measurement of CP violation in $B\to D^{\ast}\mu\nu$ decays at LHCb}

\author[a,b]{V. Dedu}
\author[a]{and A. Poluektov}
\affiliation[a]{Aix Marseille Univ, CNRS/IN2P3, CPPM, \\ Marseille, France}
\affiliation[b]{IPhU, \\ Marseille, France}

\emailAdd{vlad-george.dedu@cern.ch}
\emailAdd{anton.poluektov@cern.ch}

\usepackage{xspace} 
\usepackage{upgreek}

\def\MagUp {\mbox{\em Mag\kern -0.05em Up}\xspace}

\ifthenelse{\boolean{uprightparticles}}%
{

 \def\Pmu         {\ensuremath{\upmu}\xspace}                 
 \def\Pnu         {\ensuremath{\upnu}\xspace}                 
                  
 \def\Ppi         {\ensuremath{\uppi}\xspace}

 \def\Ptau        {\ensuremath{\uptau}\xspace}

 \def\Ppsi        {\ensuremath{\uppsi}\xspace}

 \def\PDelta      {\ensuremath{\Delta}\xspace}                 
 \def\PXi         {\ensuremath{\Xi}\xspace}                 
 \def\PLambda     {\ensuremath{\Lambda}\xspace}                 
 \def\PSigma      {\ensuremath{\Sigma}\xspace}                 
 \def\POmega      {\ensuremath{\Omega}\xspace}                 
 \def\PUpsilon    {\ensuremath{\Upsilon}\xspace}

 \def\PB      {\ensuremath{\mathrm{B}}\xspace}                 
                  
 \def\PD      {\ensuremath{\mathrm{D}}\xspace}

 \def\PJ      {\ensuremath{\mathrm{J}}\xspace}                 
 \def\PK      {\ensuremath{\mathrm{K}}\xspace}

 \def\Pb      {\ensuremath{\mathrm{b}}\xspace}                 
 \def\Pc      {\ensuremath{\mathrm{c}}\xspace}

 \def\Pi      {\ensuremath{\mathrm{i}}\xspace}

 \def\Ps      {\ensuremath{\mathrm{s}}\xspace}

 \def\thebaroffset{0.0em}
}
{

 \def\Pmu         {\ensuremath{\mu}\xspace}                 
 \def\Pnu         {\ensuremath{\nu}\xspace}                 
                  
 \def\Ppi         {\ensuremath{\pi}\xspace}

 \def\Ptau        {\ensuremath{\tau}\xspace}

 \def\Ppsi        {\ensuremath{\psi}\xspace}                 
                  
 \mathchardef\PDelta="7101
 \mathchardef\PXi="7104
 \mathchardef\PLambda="7103
 \mathchardef\PSigma="7106
 \mathchardef\POmega="710A
 \mathchardef\PUpsilon="7107
                  
 \def\PB      {\ensuremath{B}\xspace}                 
                  
 \def\PD      {\ensuremath{D}\xspace}

 \def\PJ      {\ensuremath{J}\xspace}                 
 \def\PK      {\ensuremath{K}\xspace}

 \def\Pb      {\ensuremath{b}\xspace}                 
 \def\Pc      {\ensuremath{c}\xspace}

 \def\Pi      {\ensuremath{i}\xspace}

 \def\Ps      {\ensuremath{s}\xspace}

 \def\thebaroffset{0.18em}
}
\newcommand{\offsetoverline}[2][\thebaroffset]{\kern #1\overline{\kern -#1 #2}}%

\makeatletter
\ifcase \@ptsize \relax
  \newcommand{\miniscule}{\@setfontsize\miniscule{4}{5}}
\or
  \newcommand{\miniscule}{\@setfontsize\miniscule{5}{6}}
\or
  \newcommand{\miniscule}{\@setfontsize\miniscule{5}{6}}
\fi
\makeatother

\DeclareRobustCommand{\optbar}[1]{\shortstack{{\miniscule (\rule[.5ex]{1.25em}{.18mm})}
  \\ [-.7ex] $#1$}}

\def\mup        {{\ensuremath{\Pmu^+}}\xspace}
\def\mun        {{\ensuremath{\Pmu^-}}\xspace} 

\def\taum       {{\ensuremath{\Ptau^-}}\xspace}

\def\neu        {{\ensuremath{\Pnu}}\xspace}
\def\neub       {{\ensuremath{\overline{\Pnu}}}\xspace}
\def\neum       {{\ensuremath{\neu_\mu}}\xspace}
\def\neumb      {{\ensuremath{\neub_\mu}}\xspace}
\def\neut       {{\ensuremath{\neu_\tau}}\xspace}
\def\neutb      {{\ensuremath{\neub_\tau}}\xspace}
\def\neul       {{\ensuremath{\neu_\ell}}\xspace}
\def\neulb      {{\ensuremath{\neub_\ell}}\xspace}

\def\squark    {{\ensuremath{\Ps}}\xspace}

\def\cquark    {{\ensuremath{\Pc}}\xspace}

\def\bquark    {{\ensuremath{\Pb}}\xspace}

\def\pion   {{\ensuremath{\Ppi}}\xspace}
\def\piz    {{\ensuremath{\pion^0}}\xspace}
\def\pip    {{\ensuremath{\pion^+}}\xspace}
\def\pim    {{\ensuremath{\pion^-}}\xspace}

\def\kaon    {{\ensuremath{\PK}}\xspace}

\def\KorKbar {\kern \thebaroffset\optbar{\kern -\thebaroffset \PK}{}\xspace}

\def\Kp      {{\ensuremath{\kaon^+}}\xspace}
\def\Km      {{\ensuremath{\kaon^-}}\xspace}

\def\Dbar    {{\ensuremath{\offsetoverline{\PD}}}\xspace}
\def\D       {{\ensuremath{\PD}}\xspace}
\def\Db      {{\ensuremath{\Dbar}}\xspace}
\def\DorDbar {\kern \thebaroffset\optbar{\kern -\thebaroffset \PD}\xspace}
\def\Dz      {{\ensuremath{\D^0}}\xspace}

\def\Dp      {{\ensuremath{\D^+}}\xspace}
\def\Dm      {{\ensuremath{\D^-}}\xspace}

\def\DpDm    {\ensuremath{\Dp {\kern -0.16em \Dm}}\xspace}
\def\Dstar   {{\ensuremath{\D^*}}\xspace}

\def\Dstarp  {{\ensuremath{\D^{*+}}}\xspace}
\def\Dstarm  {{\ensuremath{\D^{*-}}}\xspace}

\def\Ds      {{\ensuremath{\D^+_\squark}}\xspace}

\def\Dsm     {{\ensuremath{\D^-_\squark}}\xspace}

\def\B       {{\ensuremath{\PB}}\xspace}
\def\Bbar    {{\ensuremath{\offsetoverline{\PB}}}\xspace}

\def\BorBbar {\kern \thebaroffset\optbar{\kern -\thebaroffset \PB}\xspace}
\def\Bz      {{\ensuremath{\B^0}}\xspace}
\def\Bzb     {{\ensuremath{\Bbar{}^0}}\xspace}
\def\Bd      {{\ensuremath{\B^0}}\xspace}

\def\BdorBdbar {\kern \thebaroffset\optbar{\kern -\thebaroffset \Bd}\xspace}
\def\Bu      {{\ensuremath{\B^+}}\xspace}
\def\Bub     {{\ensuremath{\B^-}}\xspace}
\def\Bp      {{\ensuremath{\Bu}}\xspace}
\def\Bm      {{\ensuremath{\Bub}}\xspace}

\def\Bs      {{\ensuremath{\B^0_\squark}}\xspace}
\def\Bsb     {{\ensuremath{\Bbar{}^0_\squark}}\xspace}
\def\BsorBsbar {\kern \thebaroffset\optbar{\kern -\thebaroffset \Bs}\xspace}

\def\jpsi     {{\ensuremath{{\PJ\mskip -3mu/\mskip -2mu\Ppsi}}}\xspace}

\def\Y#1S{\ensuremath{\PUpsilon{(#1S)}}\xspace}

\def\LorLbar     {\kern \thebaroffset\optbar{\kern -\thebaroffset \PLambda}\xspace}

\def\BF         {{\ensuremath{\mathcal{B}}}\xspace}
\def\BR         {\BF}

\def\to                 {\ensuremath{\rightarrow}\xspace}

\def\CP                {{\ensuremath{C\!P}}\xspace}

\def\AT#1     {\ensuremath{A_{\mathrm{T}}^{#1}}\xspace}           

\def\C#1      {\ensuremath{\mathcal{C}_{#1}}\xspace}                       
\def\Cp#1     {\ensuremath{\mathcal{C}_{#1}^{'}}\xspace}                    
\def\Ceff#1   {\ensuremath{\mathcal{C}_{#1}^{\mathrm{(eff)}}}\xspace}        
\def\Cpeff#1  {\ensuremath{\mathcal{C}_{#1}^{'\mathrm{(eff)}}}\xspace}       
\def\Ope#1    {\ensuremath{\mathcal{O}_{#1}}\xspace}                       
\def\Opep#1   {\ensuremath{\mathcal{O}_{#1}^{'}}\xspace}                    

\newcommand{\nospaceunit}[1]{\ensuremath{\text{#1}}}       
\newcommand{\aunit}[1]{\ensuremath{\text{\,#1}}}       

\newcommand{\tev}{\aunit{Te\kern -0.1em V}\xspace}
\newcommand{\gev}{\aunit{Ge\kern -0.1em V}\xspace}
\newcommand{\mev}{\aunit{Me\kern -0.1em V}\xspace}
\newcommand{\kev}{\aunit{ke\kern -0.1em V}\xspace}
\newcommand{\ev}{\aunit{e\kern -0.1em V}\xspace}
 
\newcommand{\mevc}{\ensuremath{\aunit{Me\kern -0.1em V\!/}c}\xspace}
\newcommand{\gevc}{\ensuremath{\aunit{Ge\kern -0.1em V\!/}c}\xspace}
\newcommand{\mevcc}{\ensuremath{\aunit{Me\kern -0.1em V\!/}c^2}\xspace}
\newcommand{\gevcc}{\ensuremath{\aunit{Ge\kern -0.1em V\!/}c^2}\xspace}

\def\mm   {\aunit{mm}\xspace}

\def\mum  {\ensuremath{\,\upmu\nospaceunit{m}}\xspace}

\def\fb   {\ensuremath{\aunit{fb}}\xspace}
\def\invfb   {\ensuremath{\fb^{-1}}\xspace}

\def\gsim{{~\raise.15em\hbox{$>$}\kern-.85em
          \lower.35em\hbox{$\sim$}~}\xspace}
\def\lsim{{~\raise.15em\hbox{$<$}\kern-.85em
          \lower.35em\hbox{$\sim$}~}\xspace}

\def\pt         {\ensuremath{p_{\mathrm{T}}}\xspace}

\def\murad{\ensuremath{\,\upmu\nospaceunit{rad}}\xspace}

\def\rad{\aunit{rad}\xspace}

\def\tell1  {TELL1\xspace}
\def\ukl1   {UKL1\xspace}

\newcommand{\eg}{\mbox{\itshape e.g.}\xspace}
\newcommand{\ie}{\mbox{\itshape i.e.}\xspace}

\newcommand{\etc}{\mbox{\itshape etc.}\xspace}

\renewcommand{\Im}{{\rm\, Im}}

\abstract{
  Measurement of $CP$-violating observables in semileptonic 
  decays is a sensitive null-test of the Standard Model: any $CP$ violation 
  would be an unambiguous sign of New Physics effects. 
  The model-independent technique to measure parity and $CP$-odd
  observables in the $B\to D^{\ast}\mu\nu$ decays is proposed, which
  effectively cancels out parity-even terms in the decay density 
  together with the associated theory uncertainty. 
  The feasibility study is performed with pseudoexperiments, and the
  sensitivity at the LHCb experiment is estimated. Finally, the most
  significant systematic effects and the data-driven ways to control them are considered. 
}

\begin{document} 
\maketitle
\flushbottom

\section{Introduction}
\label{sec:introduction}

An important property of the Standard Model of electroweak interactions (SM) is lepton flavour universality (LFU): the couplings of the three generations of leptons to the electroweak bosons are exactly the same. This property can be violated in a number of New Physics (NP) scenarios. Numerous tests of LFU have been performed by the experiments studying the decays of $\B$ hadrons by comparing the rates of the $\B$ hadron decays mediated by the neutral current involving $\ell^+\ell^-$ pair (where $\ell=e, \mu, \tau$)~\cite{BaBar:2012mrf,Belle:2016fev,BELLE:2019xld,Belle:2019oag,LHCb:2014vgu,LHCb:2017avl,LHCb:2019hip,LHCb:2019efc,LHCb:2021trn,LHCb:2021lvy,LHCb:2022qnv} or in the charged current decays with $\ell\neul$
combination~\cite{BaBar:2012obs,BaBar:2013mob,Belle:2015qfa,Belle:2016dyj,Belle:2017ilt,Belle:2019rba,LHCb:2017smo,LHCb:2015gmp,LHCb:2017rln}. 

In searches for NP effects in charged currents, in addition to the ratios of branching fractions ($R(\Dstar)$, $R(D)$, \etc) probing LFU, other observables have been suggested that could potentially help distinguish between various NP models if significant deviations from the SM predictions are found. These include various polarisation observables (such as longitudinal $\Dstar$ polarisation $F_L$~\cite{Belle:2019ewo,Belle:2023bwv}, forward-backward asymmetry $A_{FB}$~\cite{Bobeth:2021lya,Bhattacharya:2022bdk,Belle:2023bwv}, $\tau$ lepton polarisation accessible via multibody $\tau$ decays~\cite{Belle:2016dyj,Belle:2017ilt}), and generally the parameters accessible via the analyses of full angular distributions in semileptonic decays~\cite{Becirevic:2019tpx,Duraisamy:2013pia,Huang:2021fuc,Hill:2019zja}. 

Searches for $CP$ violation in semileptonic decays have been proposed as a promising way to search for NP effects that is complementary to the measurement of other observables~\cite{Duraisamy:2013pia,Bhattacharya:2019olg}. Such analysis is basically a {\it null-test} of the SM. Since the necessary condition for the $CP$ violation is the presence of two amplitudes with a nonzero weak phase difference, while in the SM the semileptonic decays proceed via a single tree-level transition, any sign of $CP$ violation in such decays would provide unambiguous evidence of NP. Another condition of nonzero $CP$ violation is that the two interfering amplitudes, SM and NP ones, have different strong phases: as a result, not all NP scenarios can give rise to $CP$-violating effects. 

$CP$-violating contributions are possible in the $\Bzb\to \Dstarp\ell^-\neulb$ decays\footnote{Here and in the following, the addition of charge conjugate modes is implied unless explicitly mentioned otherwise.} if the full angular distribution is measured. Another mechanism is possible in the decays with higher $D$ meson excitations in the final state ($\Bm\to \D^{**0}(\to\Dstarp\pim)\mun\neumb$). In that case, due to the interference of two or more overlapping $\D^{**0}$ states with different quantum numbers, the strong phase difference appears from resonant behaviour, and $CP$ asymmetry remains even if some of the kinematic degrees of freedom are integrated out~\cite{Aloni:2018ipm}. 

Semileptonic decays with $\tau$ leptons in the final state are usually considered as those that are potentially the most affected by NP. However, from the experimental point of view, these final states are much more difficult to deal with, since the secondary decay of $\tau$ lepton contains additional one or two unreconstructable neutrinos. It is thus logical to start the experimental programme of $CP$-violation studies from the $\Bzb\to \Dstarp\mun\neumb$ decays. 

Experimental studies of semileptonic decays, especially in proton collisions, are complicated by the unreconstructable neutrinos in the final state. As a result, these measurements are affected by significant backgrounds, and experimental resolution effects play a major role in the study of the angular distribution. This paper discusses the feasibility of the precision measurement of $CP$ violation in $\Bzb\to \Dstarp\mun\neumb$ decays in proton collisions at the LHCb experiment~\cite{Alves:2008zz}. 

The paper is organised as follows. Section~\ref{sec:pheno} introduces the mechanisms of parity and $CP$ violation in the $\bquark\to\cquark\ell^{-}\neulb$ transitions. Section~\ref{sec:simulation} describes the Monte-Carlo simulation procedures used in the study. Reconstruction of missing neutrino specific to LHCb that is crucial for the understanding of possible systematic biases in $P$-odd observables is described in Section~\ref{sec:reconstruction}. Section~\ref{sec:binned_fit} presents the model-independent binned fit technique proposed to study the $P$-odd part of the decay density, and the estimate of statistical sensitivity of this technique is reported in Section~\ref{sec:stat}. The most important sources of systematic uncertainty that could mimic the NP signal (parity and $CP$ violation in backgrounds, various instrumentation effects) and the ways to control them in a data-driven way are discussed in Section~\ref{sec:systematics}. We conclude in Section~\ref{sec:conclusion}. 

\section{Phenomenology of $CP$ violation in $\Bzb\to \Dstarp\ell^-\neulb$ decays}

\label{sec:pheno}

\def \cA {{\cal A}}
\def \mAp{\mathcal{A}_+}
\def \mAm{\mathcal{A}_-}
\def \mAn{\mathcal{A}_{0}}
\def \mAt{\mathcal{A}_{t}}
\def \mAperp{\mathcal{A}_\perp}
\def \mApar{\mathcal{A}_\parallel}
\def \mAperpT{\mathcal{A}_{\perp, T}}
\def \mAparT{\mathcal{A}_{\parallel, T}}
\def \mA0T{\mathcal{A}_{0, T}}

The decay $\Bzb\to \Dstarp\mun\neumb$ is fully described by four kinematic variables: the invariant mass squared of the lepton system $q^2=m^2(\mun\neumb)$, two helicity angles $\cos\theta_D$ and $\cos\theta_{\ell}$, and the azimuthal angle $\chi$ between the $\Dstarp\to \Dz\pip$ and $\mun\neumb$ decay planes in the $\Bzb$ rest frame ($\chi\in [-\pi, \pi]$). The definition of kinematic variables from Ref.~\cite{Becirevic:2019tpx} (Section~2.3) is used here. The fully differential decay density as a function of these variables can be split into parity-even and parity-odd parts, 
\begin{equation}
    d\Gamma = (P_{\rm even} + P_{\rm odd})\;dq^2\;d\cos\theta_{D}\;d\cos\theta_{\ell}\;d\chi, 
\end{equation}
where $P_{\rm even}$ is the even and $P_{\rm odd}$ is the odd function of $\chi$, respectively. In the SM, $P_{\rm odd}\equiv 0$, while it can have non-zero terms in two NP cases: the contribution of right-handed vector current (RH, in which case $P_{\rm odd}$ contains the terms proportional to $\sin\chi$ and $\sin 2\chi$), and the interference of the pseudoscalar and tensor currents (PT, with only the contribution proportional to $\sin\chi$). The angular functions associated with these two NP cases are given in Table~\ref{tab:cpv_terms_couplings} extracted from Ref.~\cite{Bhattacharya:2019olg}\footnote{The table in Ref.~\cite{Bhattacharya:2019olg} contains additional angular term proportional to $ {\Im}(\mAn \mApar^*)$. However, as follows from the definitions of the $\mAn$ and $\mApar$ amplitudes, this term should not result in $CP$ violation even for nonzero NP couplings. This is also supported by the independent derivation of the amplitudes in Ref.~\cite{Ligeti:2016npd}. We are grateful to the journal referee for pointing this out.}. The explicit expressions for the full decay density can be found, \eg, in Refs.~\cite{Becirevic:2019tpx, Ligeti:2016npd}. The $P_{\rm odd}$ part of the decay density corresponds to combination of $I_7$, $I_8$, and $I_9$ terms in Ref.~\cite{Becirevic:2019tpx}, Eq.~(17).

\begin{table}
\caption{Unsuppressed \CP-violating terms in the angular distribution and their NP couplings}
\label{tab:cpv_terms_couplings}
\begin{center}
\begin{tabular}{llr}
\toprule
Amplitude term & Coupling & Angular function \\
\midrule
${\Im}(\mAperp \mAn^*)$            &     $\Im[(1+g_L+g_R)(1+g_L-g_R)^*]$       &      $- \sqrt{2} \sin 2\theta_\ell \sin 2\theta_D \sin\chi$               \\
$ {\Im}(\mApar \mAperp^*)$          &    $\Im[(1+g_L - g_R)(1+g_L+g_R)^*]$       &     $ 2 \sin^2 \theta_\ell \sin^2 \theta_D \sin 2\chi$                \\
$ {\Im}(\cA_{SP} \mAperpT^*)$        &   $\Im(g_Pg_T^*)$  &       $- 8\sqrt{2} \sin\theta_\ell \sin 2\theta_D\sin\chi$              \\
\bottomrule
\end{tabular}
\end{center}
\end{table}

The full angular analysis such as the one suggested in Ref.~\cite{Hill:2019zja} is sensitive, in particular, to $CP$-violating terms in the decay density. However, a dedicated analysis is likely to be needed to reach the optimal precision for $CP$-violating observables. It could utilise the symmetry properties of the decay density to completely cancel out the parity-even part of the distribution to reduce the associated uncertainty, \eg, due to $q^2$ formfactor dependence. On the other hand, good control of parity-odd effects in the backgrounds and detector response is specific to such analysis and should be addressed in more detail. Finally, the analysis procedure for the dedicated $CP$ violation measurement could be simplified compared to the full four-dimensional decay density fit. 

The parity-odd component $P_{\rm odd}$ of the decay density can, in general, be either parity ($P$) or charge-parity ($CP$) violating, depending on how it is related for the $\Bzb$ and charge conjugate $\Bz$ decays. Parity transformation corresponds to the sign flip $\chi\to -\chi$, thus to the sign flip for the $P_{\rm odd}$ term if the same definition of the angle $\chi$ is used for the process and its anti-process. Then, charge conjugation corresponds to the sign flip of the weak phase difference, and, since $P_{\rm odd}$ is proportional to the imaginary part of the NP coupling, it corresponds again to the flip of the $P_{\rm odd}$ sign. Therefore, in the case of $CP$-odd processes, the sign of the term $P_{\rm odd}$ should be \emph{the same} for the process and anti-process, while for $P$-odd processes the sign \emph{changes}. 

The amplitude terms $\mAperp \mAn^*$ and $\mApar \mAperp^*$ 
which appear in Table~\ref{tab:cpv_terms_couplings} are non-zero in the SM. Thus, if they had non-zero strong phase difference with respect to the rest of the amplitude, one would observe parity violation in the decay density with $P_{\rm odd}$ having the opposite sign for the $\Bzb$ and $\Bz$ decays. It is expected, however, that the strong phase is the same for all the amplitudes in the $\Bzb\to \Dstarp\mun\neumb$ decay~\cite{Bhattacharya:2019olg}. Therefore, testing if the $P_{\rm odd}$ component is the same for the $\Bzb$ and $\Bz$ decay can serve as a good consistency check in the experiment that is independent of the presence of NP. 

\section{Simulation of $\Bzb\to \Dstarp\mun\neumb$ decays}
\label{sec:simulation}

The study of the feasibility of $CP$ violation measurement in $\Bzb\to \Dstarp\mun\neumb$ decays presented in this publication is based on the simplified Monte Carlo simulation. The kinematic parameters of the $\Bzb\to \Dstarp\mun\neumb$ decays are generated according to the amplitude model from Ref.~\cite{Ligeti:2016npd}. Since the performance of the method is not expected to be strongly dependent on the precise parametrisation of the formfactors, leading-order formfactor approximation is used for simplicity. The NP couplings $g_{R,P,T}$ varied in the current study are related to the quark and lepton couplings from Ref.~\cite{Ligeti:2016npd} as 
\begin{equation}
    \begin{split}
        g_P & = (\alpha^S_L-\alpha^S_R)\beta^S_L r^2_S, \\
        g_R & = \alpha^V_R\beta^S_L r^2_V, \\
        g_T & = \alpha^T_R\beta^S_L r^2_T, \\
    \end{split}
\end{equation}
where $\alpha$ and $\beta$ are the NP couplings of quark and lepton currents, respectively, and $r_{S,V,T}=m_W/\Lambda_{S,V,T}$ defines the NP mass scale. 

To obtain the laboratory-frame momenta of the decay products, initial $\Bzb$ mesons are generated with the transverse momentum $\pt$ and pseudorapidity $\eta$ distributions from FONNL calculations~\cite{Cacciari:1998it} for 13\tev $pp$ collisions in a similar way as in the {\tt RapidSim} framework~\cite{Cowan:2016tnm}. The decay position of the $\Bzb$ and the momenta of the final state tracks are then generated by taking the exponential $\Bzb$ lifetime distribution with the known average lifetime, the spherically-symmetric orientation of the decay and using the values of the kinematic parameters $q^2$, $\cos\theta_{D}$, $\cos\theta_{\ell}$ and $\chi$ generated at the previous step. 

To simulate the reconstruction and selection requirements in real experimental conditions, only the events that satisfy certain criteria are retained. Each charged particle is required to have pseudorapidity in the range $2<\eta<5$, momentum in the range $3<p<150\gev$, and transverse momentum $\pt>300\mev$. To simulate the trigger selection, at least one particle of the $\Dz$ decay products is required to have $\pt>800\mev$. Finally, the $\Bzb$ decay vertex is required to be displaced from its production point by more than 5\mm. 

For the retained events, the impact parameter of each final-state track is smeared according to the parametrised single-track impact parameter resolution~\cite{Aaij:2014zzy}, 
\begin{equation}
  \sigma_{\rm IP} = 11.6\mum + \frac{23.4\mum}{\pt[\gev]}. 
\end{equation}
The $\Dstarp\mun$ vertex position is then fitted back using a simple least-squares fit. The fitted position of the vertex and the momenta of four charged final-state particles are used in the kinematic reconstruction of the decay parameters as described in the next Section. 
Since the precision of the decay parameters discussed in the next Section is by far dominated by the precision of the reconstruction of the \B vertex, finite precision of the track momentum reconstruction is not simulated. 

\section{Reconstruction of decay parameters}
\label{sec:reconstruction}

Using the topological information from the position of the primary vertex (the origin vertex of the $\B$ meson) and the $\B$ meson decay vertex, one can obtain all the kinematic parameters of the $\Bzb\to \Dstarp\mun\neumb$ decay with the missing neutrino up to a quadratic ambiguity~\cite{Dambach:2006ha}. The absolute value $p_B$ of the momentum of the $B$ meson in the laboratory frame can be estimated from the observable quantities of the $\Bzb\to \Dstarp\mun\neumb$ candidate as 
\begin{equation}
  p_{B} = \frac{\left(m_{B}^{2} + m^2_{\Dstar\mu} \right)p_{\Dstar\mu}\cos\theta \pm E_{\Dstar\mu}
  \sqrt{(m_B^2 - m^2_{\Dstar\mu})^2 - 4 m_B^2 p^2_{\Dstar\mu}\sin^2\theta}
  }{2(m_{\Dstar\mu}^{2} + p_{\Dstar\mu}^{2}\sin^2\theta)}, 
  \label{eq:b_momentum}
\end{equation}
where $m_{\Dstar\mu}$, $p_{\Dstar\mu}$ and $E_{\Dstar\mu}$ are the invariant mass, momentum, and energy of the $\Dstarp\mun$ combination in the laboratory frame, $m_B$ is the $\B$ meson mass, and $\theta$ is the angle between the direction of the $B$ meson (which can be reconstructed from the positions of the $B$ vertex and the primary vertex of the $pp$ interaction) and the $\Dstarp\mun$ combination. 

The two solutions of the Eq.~\ref{eq:b_momentum} are referred to as ``$+$'' and ``$-$''. Due to the finite precision of the $B$ vertex reconstruction, in some cases the expression under the square root in Eq.~\ref{eq:b_momentum} is negative. In such cases, the closest ``physical'' solution is chosen by assigning the square root to be equal to zero. In the following studies, we will compare the performance of the two solutions, and the third approach when the average of the two is taken (\ie the case when the square root in Eq.~\ref{eq:b_momentum} is always taken to be zero; this solution is referred to as ``$\rm avg$''). 

\begin{figure}
    \centering
    \includegraphics[width=0.48\textwidth]{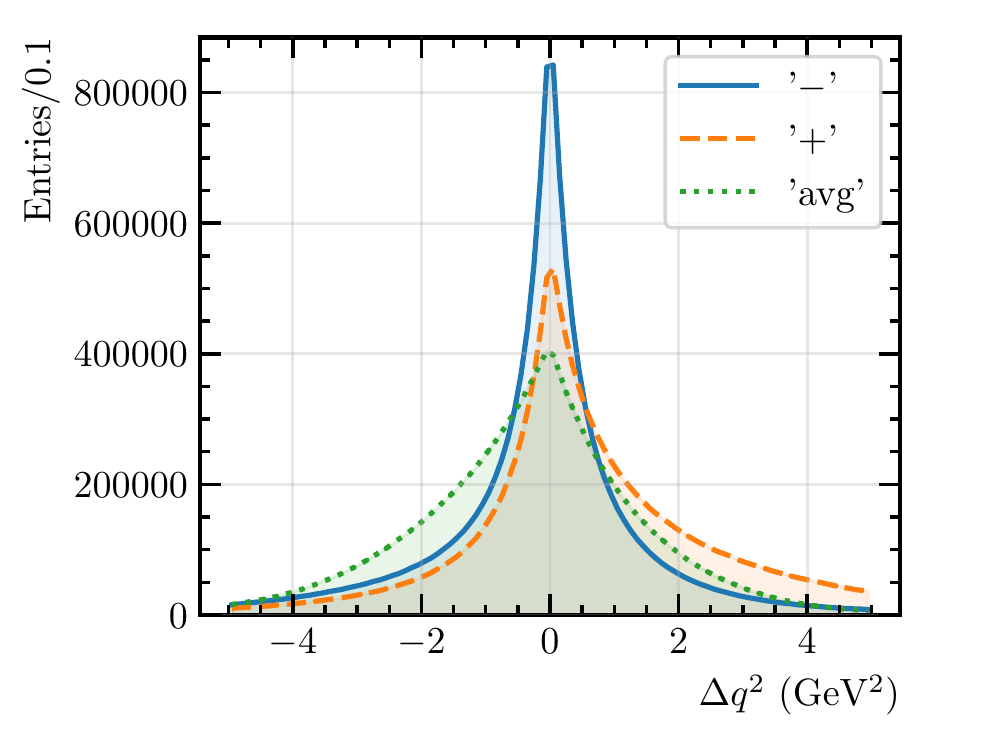}
    \put(-155, 130){(a)}
    \includegraphics[width=0.48\textwidth]{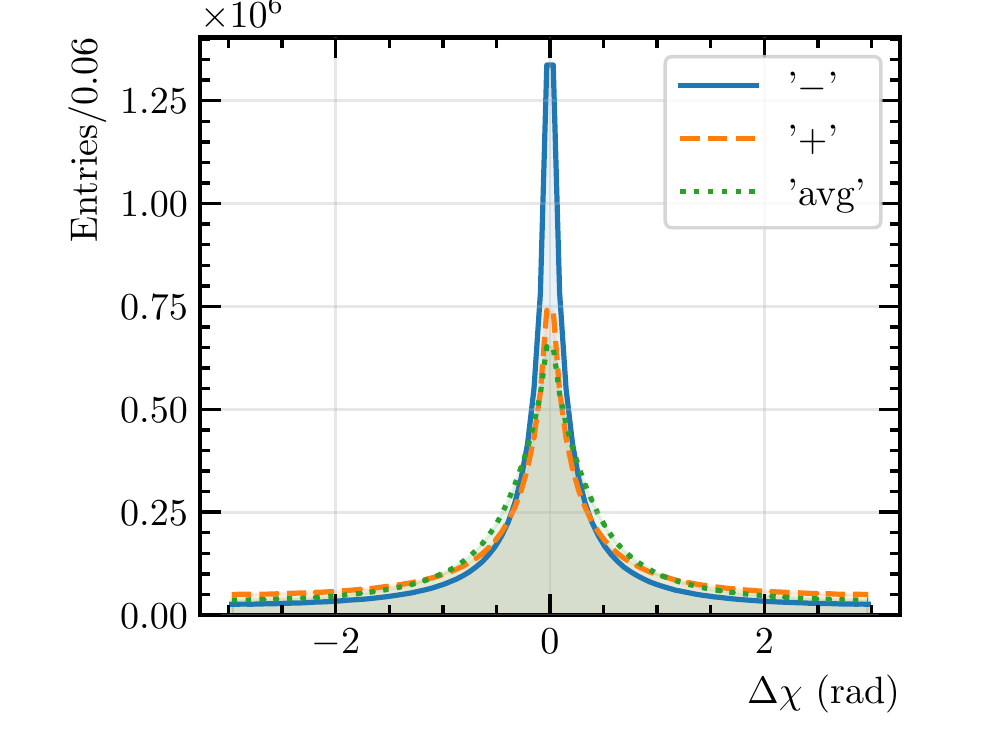}
    \put(-155, 130){(b)}
    
    \includegraphics[width=0.48\textwidth]{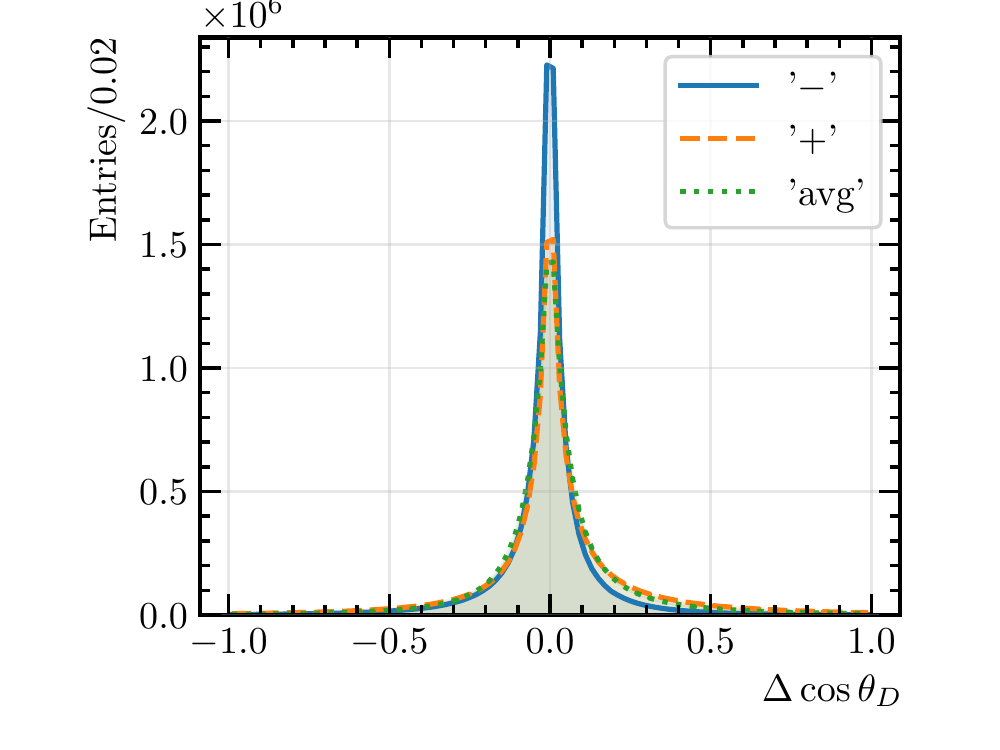}
    \put(-155, 130){(c)}
    \includegraphics[width=0.48\textwidth]{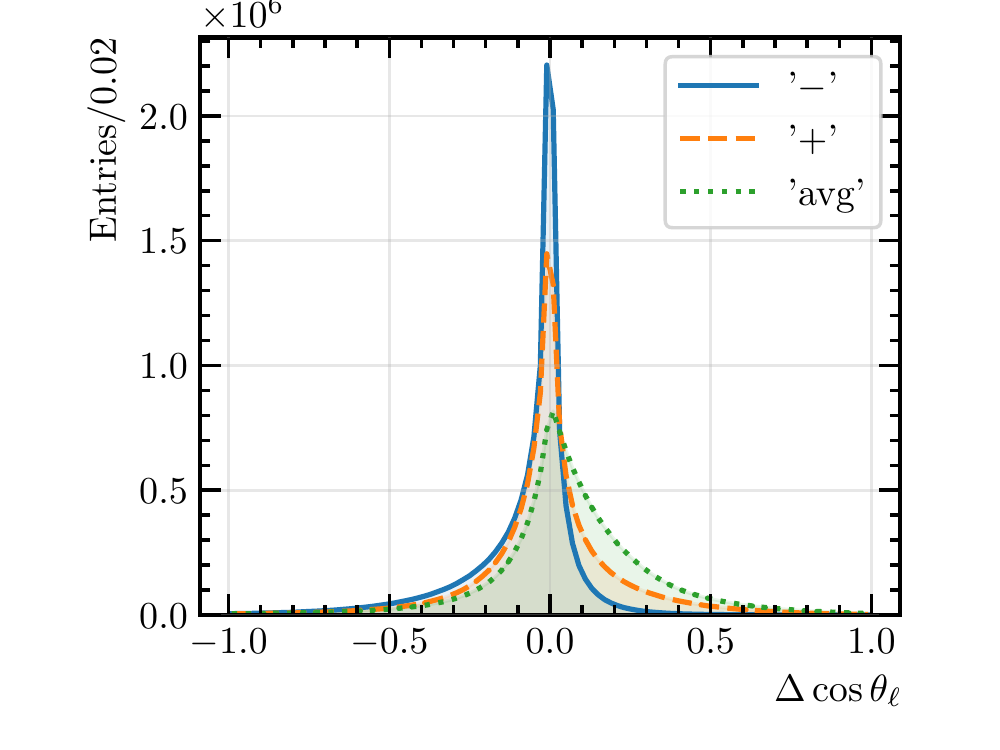}
    \put(-155, 130){(d)}

    \caption{Residual distributions for the different solutions of the reconstructed $\Bzb\to \Dstarp\mun\neumb$ decay parameters (a) $q^2$, (b) $\chi$, (c) $\cos\theta_D$ and (d) $\cos\theta_{\ell}$. }
    \label{fig:decay_param_resid}
\end{figure}

Once $p_B$ is determined, the 3-momentum of the $\Bzb$ meson $\vec{p}_B$ can be calculated by multiplying $p_B$ by the $\Bzb$ flight direction using the displacement of its decay vertex $\vec{n}$. The kinematic variables $q^2$, $\cos\theta_{D}$, $\cos\theta_{\ell}$ and $\chi$ are then calculated. Figure~\ref{fig:decay_param_resid} shows the comparison of the residual distributions of the four parameters for the simulated $\Bzb\to \Dstarp\mun\neumb$ sample using three solutions. It is clear that the solution ``$-$'' offers the best resolution. Since the distributions are non-Gaussian, it is hard to quantify the resolutions of the reconstructed quantities. In the following Section, the performance of the three solutions will be compared based on the statistical precision of NP couplings. 
 
To study the possible systematic effects that could lead to fake parity-odd terms in the measured decay density, it is convenient to express 
the value of $\sin\chi$ using the momenta of the reconstructable decay products in the $B$ rest frame as
\begin{equation}
    \sin\chi = \left(\frac{\vec{p}^{\;'}_{\mu}\times \vec{p}^{\;'}_{D^*}}{|\vec{p}^{\;'}_{\mu}\times \vec{p}^{\;'}_{D^*}|}\times \frac{\vec{p}^{\;'}_{\pi}\times \vec{p}^{\;'}_{D^*}}{|\vec{p}^{\;'}_{\pi}\times \vec{p}^{\;'}_{D^*}|}, \frac{\vec{p}^{\;'}_{D^*}}{|\vec{p}^{\;'}_{D^*}|}\right) = 
    -\frac{(\vec{p}^{\;'}_{\pi}, \vec{p}^{\;'}_{\mu}, \vec{p}^{\;'}_{D})}{|\vec{p}^{\;'}_{\mu}\times \vec{p}^{\;'}_{D^*}|\,|\vec{p}^{\;'}_{\pi}\times \vec{p}^{\;'}_{D^*}|}|\vec{p}^{\;'}_{D^*}|. 
\end{equation}
After boosting to the laboratory frame using the estimation of the $B$ momentum $\vec{p}_B$, $\sin\chi$ can be represented as a combination of four triple products of the momenta defined in the laboratory frame:  
\begin{equation}
  \sin\chi = 
    S_1\cdot (\vec{p}_{\pi}, \vec{p}_{\mu}, \vec{p}_D) + 
    S_2\cdot (\vec{p}_{B}, \vec{p}_{\mu}, \vec{p}_D) + 
    S_3\cdot (\vec{p}_{\pi}, \vec{p}_{B}, \vec{p}_D) + 
    S_4\cdot (\vec{p}_{\pi}, \vec{p}_{\mu}, \vec{p}_{B}), 
  \label{eq:sinchi_lab}
\end{equation}
where $S_i$ are $P$-even functions of decay kinematics (\eg of $|p_i|$ or $(\vec{p}_i,\vec{p}_j)$), while the triple products $(p_i,p_j,p_k)$ are $P$-odd. 

It is important to note that the $\sin\chi$ value reconstructed as in Eq.~\ref{eq:sinchi_lab}, even though estimated from the topological information, is still purely $P$-odd as is the true one. It can thus be used as a per-event weight to cancel out the $P$-even contribution in data. However, certain reconstruction effects can result in the $\sin\chi$ variable not being exactly $P$-odd, in which case one would observe a fake $P$-odd signal in the distribution that is purely SM-like ($P$-even). To analyse which detector effects can produce a fake $P$-odd signal, it is useful to further expand the expression~(\ref{eq:sinchi_lab}) in terms of spherical coordinates where the polar axis is given by the direction of the beams: 
\begin{equation}
    \sin\chi = \sum\limits_{i\neq j}S_{ij}\cdot \sin(\phi_i-\phi_j), 
    \label{eq:sinchi_xy}
\end{equation}
where $i,j$ are the indices corresponding to the momenta of the $B$ and final state tracks, $\phi_i$ are the azimuthal angles of the direction of particle $i$, and $S_{ij}$ are $P$-even functions of decay kinematics. Since the production of $\B$ mesons is uniform in $\phi$ up to high precision, the fake $P$-odd signal can only be generated by the detector effects that produce nonzero $\sin(\phi_i-\phi_j)$ terms after averaging over $\phi$. Such effects will be discussed in Section~\ref{sec:systematics}. 

\section{Binned asymmetry fit}
\label{sec:binned_fit}

The parity-odd part of the decay density, $P_{\rm odd}$, is the sum of terms proportional to $\sin\chi$ and $\sin 2\chi$: 
\begin{equation}
    P_{\rm odd}(q^2, \theta_D, \theta_{\ell}, \chi) = 
    P^{(1)}_{\rm odd}(q^2, \theta_D, \theta_{\ell})\sin\chi + 
    P^{(2)}_{\rm odd}(q^2, \theta_D, \theta_{\ell})\sin 2\chi. 
    \label{eq:p_odd_composition}
\end{equation}
These terms can be obtained by integrating the full decay density $P_{\rm tot}=P_{\rm even}+P_{\rm odd}$ with the weights $\sin\chi$ and $\sin 2\chi$, respectively: 
\begin{equation}
  \begin{split}
    P^{(1)}_{\rm odd}(q^2, \theta_D, \theta_{\ell}) = & \,\frac{1}{\pi}\int\limits_{-\pi}^{\pi}P_{\rm tot}(q^2, \theta_{D}, \theta_{\ell}, \chi)\sin\chi\;d\chi, \\
    P^{(2)}_{\rm odd}(q^2, \theta_D, \theta_{\ell}) = & \,\frac{1}{\pi}\int\limits_{-\pi}^{\pi}P_{\rm tot}(q^2, \theta_{D}, \theta_{\ell}, \chi)\sin 2\chi\;d\chi. 
  \end{split}
  \label{eq:p_odd_parts}
\end{equation}
As a result, the SM-dominated $P_{\rm even}$ contribution is cancelled out in a model-independent way. 

Assuming that the NP couplings are small compared to the SM one (\eg, in the limit $g_R\ll 1$ and $g_Pg_T^*\ll 1$), the two terms in Eq.~\ref{eq:p_odd_composition} are\footnote{The second-order term proportional to $\Im(g_Lg_R^*)$ is ignored here for simplicity}
\begin{equation}
  \begin{split}
    P^{(1)}_{\rm odd}(q^2, \theta_D, \theta_{\ell}) = & \Im(g_R)F^{(1)}_{RH}(q^2, \theta_D, \theta_{\ell}) + \Im(g_P g_T^*)F^{(1)}_{PT}(q^2, \theta_D, \theta_{\ell}), \\
    P^{(2)}_{\rm odd}(q^2, \theta_D, \theta_{\ell}) = & \Im(g_R)F^{(2)}_{RH}(q^2, \theta_D, \theta_{\ell}). \\
  \end{split}
  \label{eq:p_odd_expr}
\end{equation}
Right-handed current gives rise to the $P$-odd signal in both $\sin\chi$ and $\sin 2\chi$ terms, while the pseudoscalar-tensor interference only results in non-zero term with $\sin\chi$ (see Table~\ref{tab:cpv_terms_couplings}). The system of equations~\ref{eq:p_odd_expr} can be solved in the experiment to obtain the constraints on the P-odd combinations of NP couplings $\Im(g_R)$ and $\Im(g_P g_T^*)$ using the terms $P_{\rm odd}^{(1,2)}(q^2, \theta_D, \theta_{\ell})$ measured in data and the functions $F_{RH,PT}^{(1,2)}(q^2, \theta_D, \theta_{\ell})$ that can be obtained from the simulated samples. This way, one can avoid dealing with the full angular fit to obtain the parity-odd observables, and consequently remove the uncertainty related to the description of the (dominant) parity-even part of the decay density. 

In a real experiment, one has to deal with scattered data rather than with decay densities. In addition, the decay parameters are not the true ones, but rather the approximated values (see Section~\ref{sec:reconstruction}) which are affected by the experimental resolution and non-uniform acceptance. In practice, it is convenient to deal with binned asymmetries $A_i$ which are related to the density terms $P(q^2, \cos\theta_D, \cos\theta_{\ell})$ by a linear transformation (since multiplication by the efficiency, convolution with the detector resolution, and calculation of the integral over the bin area are all linear operations). As a result, a linear template fit using the binned version of Eq.~\ref{eq:p_odd_expr} can be used to extract the NP couplings. 

To be specific, the binned asymmetries in the $i$-th bin are defined as 
\begin{equation}
    A^{(1)}_i = \frac{N_{\rm bins}}{N_{\rm signal}}\sum\limits_{n=1}^{N_i} \sin\chi_n, \;\;\;
    A^{(2)}_i = \frac{N_{\rm bins}}{N_{\rm signal}}\sum\limits_{n=1}^{N_i} \sin 2\chi_n, \;\;\;
    \label{eq:binned_asym}
\end{equation}
where the summation is performed over all the events that belong to the $i$-th bin in the $q^2, \cos\theta_D, \cos\theta_{\ell}$ region, $N_{\rm signal}=\sum_i N_i$ is the total number of signal events in the sample and $N_{\rm bins}$ is the number of bins ($1\leq i \leq N_{\rm bins}$). The normalisation term $N_{\rm bins}/N_{\rm signal}$ is chosen such that 1) the magnitudes of the asymmetries are independent of the number of events in the sample, and 2) the asymmetries are independent of the parity-even part of the density (as would be the case if, \eg, the term $1/N_i$ was used for the normalisation). 

\begin{figure}
    \centering
    \includegraphics[width=0.48\textwidth]{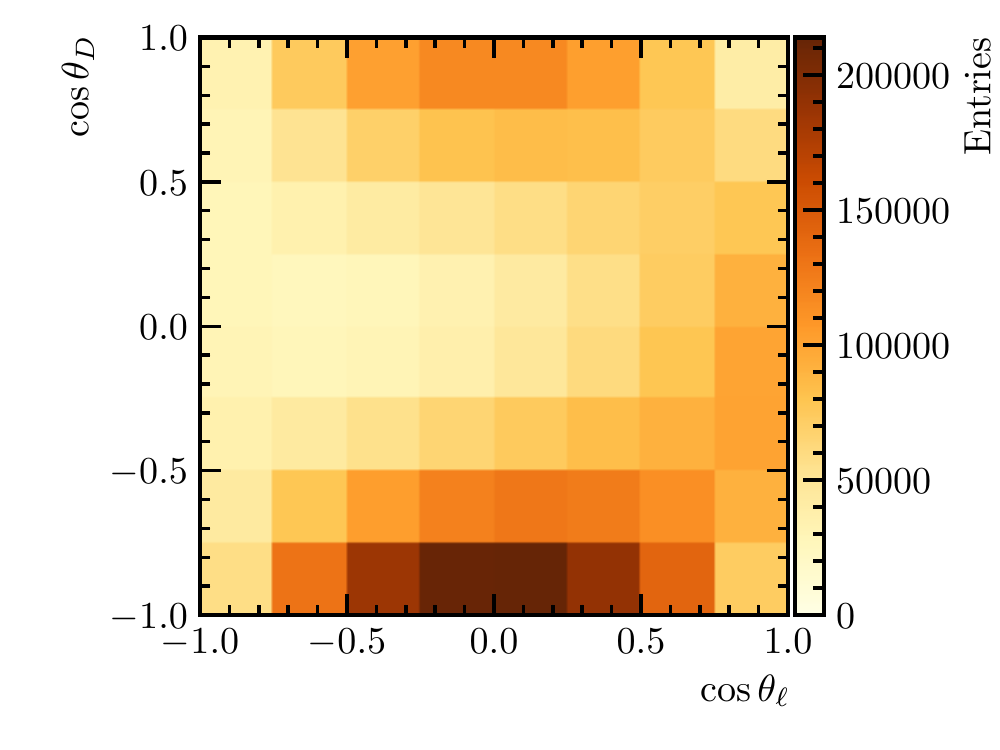}
    \put(-159, 131){\colorbox{white}{(a)}}

    \includegraphics[width=0.48\textwidth]{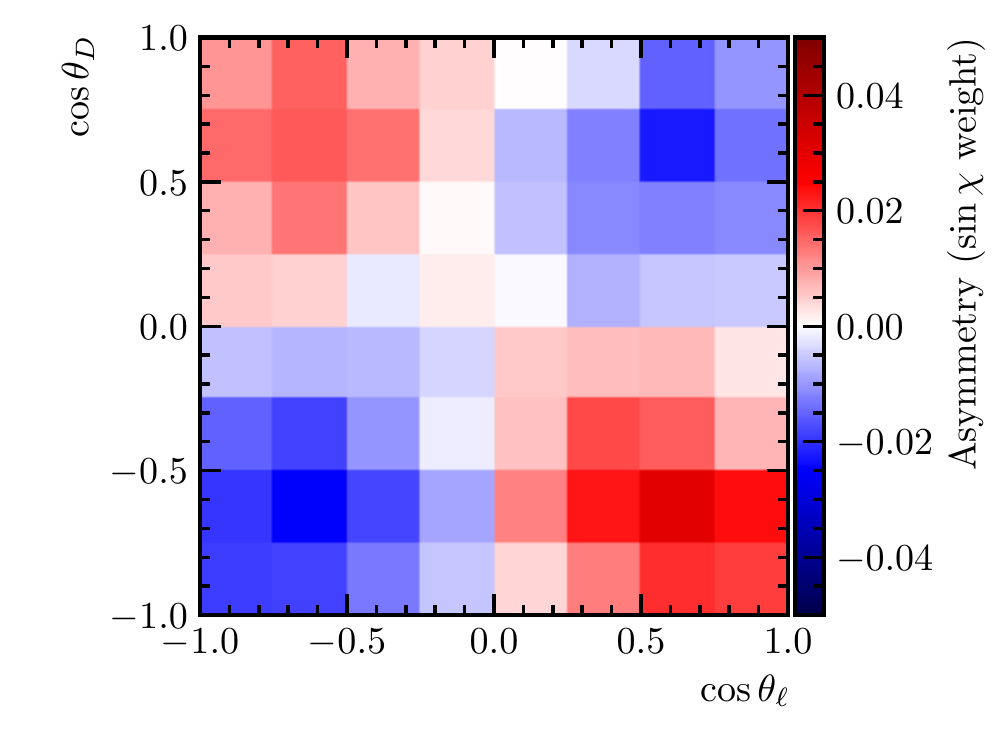}
    \put(-159, 131){\colorbox{white}{(b)}}
    \includegraphics[width=0.48\textwidth]{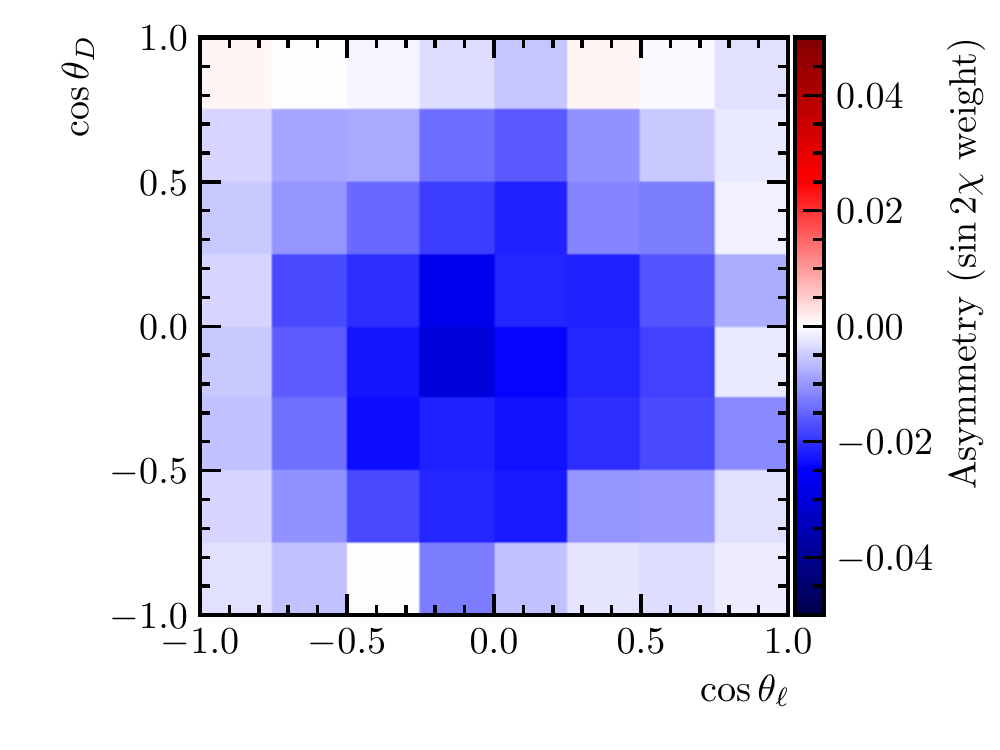}
    \put(-159, 131){\colorbox{white}{(c)}}

    \includegraphics[width=0.48\textwidth]{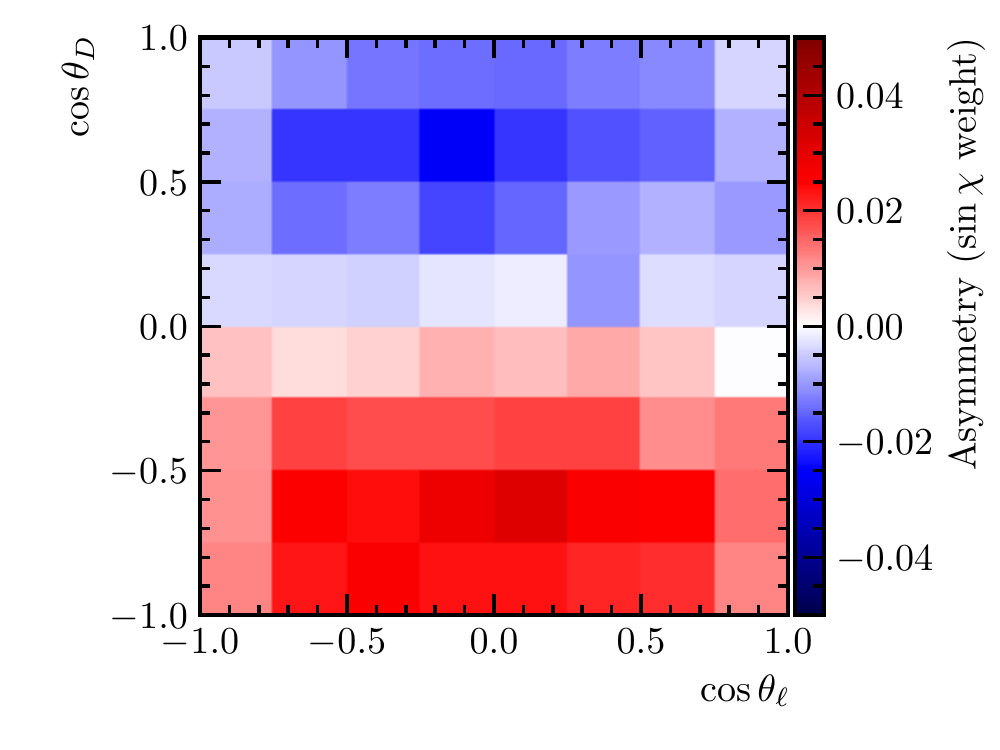}
    \put(-159, 131){\colorbox{white}{(d)}}
    \includegraphics[width=0.48\textwidth]{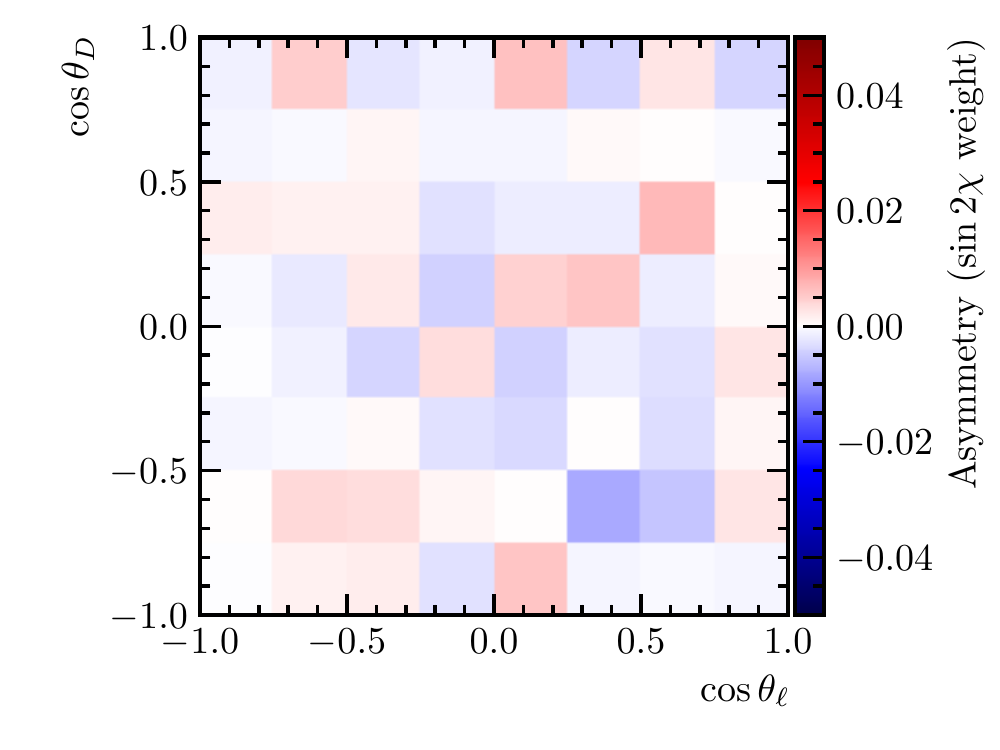}
    \put(-159, 131){\colorbox{white}{(e)}}

    \caption{(a) Binned density and binned asymmetries of (b,d) $\sin\chi$ and (c,e) $\sin 2\chi$
    terms in $(\cos\theta_D, \cos\theta_{\ell})$ bins integrated over $q^2$ for the (b,c) contribution of right-handed current with $\Im(g_R)=0.1$ and (d,e) interference of pseudoscalar and tensor currents with $\Im(g_P g_T^*)=0.02$. Truth-level angles are used. }
    \label{fig:asym_rh_pt}
\end{figure}

\begin{figure}
    \centering
    \includegraphics[width=0.48\textwidth]{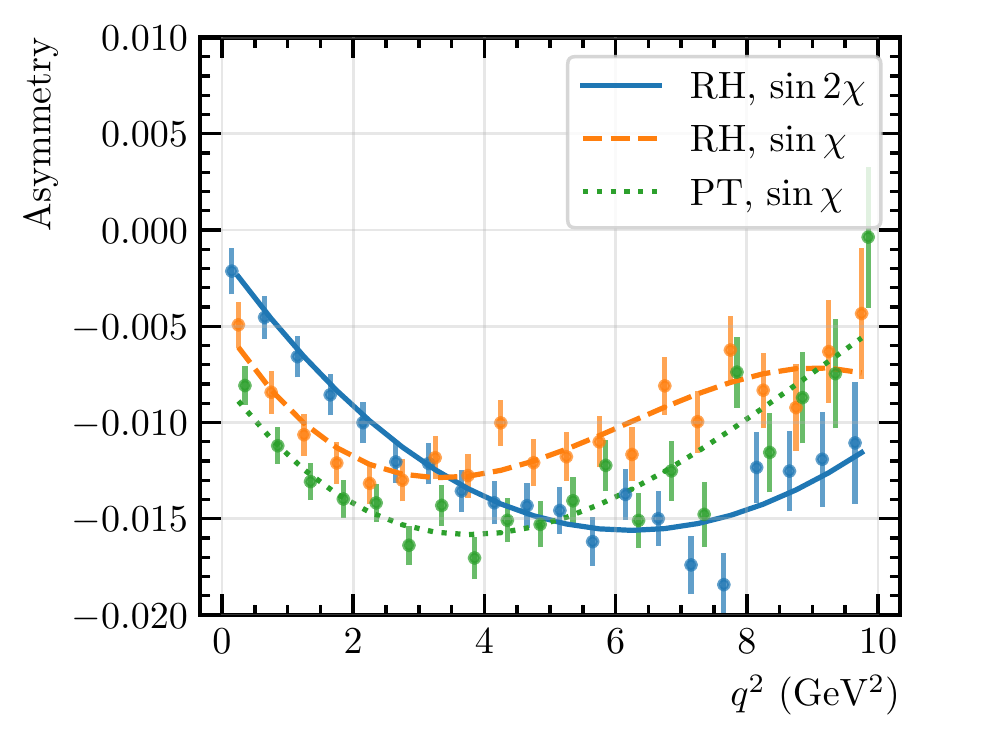}

    \caption{$q^2$ dependence of the asymmetry terms. Only the $\cos\theta_D>0$, $\cos\theta_{\ell}>0$ part of the phase space is taken to avoid cancellation of the asymmetry for the $\sin\chi$ terms. Truth-level angles are used. }
    \label{fig:asym_q2}
\end{figure}

Figure~\ref{fig:asym_rh_pt} demonstrates the binned asymmetries obtained from the samples with the admixtures of right-handed current with $\Im(g_R)=0.1$ and interference of pseudoscalar and tensor current with $\Im(g_P g_T^*)=0.02$. True angles are used in the calculation of templates. 
The asymmetries are obtained by integrating over the allowed $q^2$ range. Figure~\ref{fig:asym_q2} shows that it does not lead to cancellation of the asymmetry, since it has the same sign in the full $q^2$ range. In what follows, we will thus consider only the 2D binned asymmetries in the $\cos\theta_{D,\ell}$ bins without splitting in $q^2$. In a real measurement, one might still consider splitting the sample in $q^2$ as an additional consistency check. 

The observed binned asymmetries~(\ref{eq:binned_asym}) can now be used to extract the values of the NP couplings $\Im(g_R)$ and $\Im(g_P g_T^*)$. Since the values of the asymmetry in each bin are sums of a large number of independent terms, their fluctuations are Gaussian to a good precision (if the bins are not too small) and a simple $\chi^2$-like function could be used to fit the binned asymmetry to the model represented by the binned version of Eq.~\ref{eq:p_odd_expr}. One should keep in mind, however, that the $\sin\chi$ and $\sin 2\chi$ asymmetry terms in the same bin $i$ are obtained from the same data set and are thus correlated. The following function can thus be used to take this correlation into account: 
\begin{equation}
    \chi^2_{\rm corr} = \sum\limits_{i}\sum\limits_{a,b=1,2}\Delta A_i^{(a)}\;\left(\Sigma^{-1}_i\right)^{(ab)}\;\Delta A_i^{(b)}. 
    \label{eq:chi2_fit}
\end{equation}
Here the indices $a,b$, which can take the values 1 or 2, refer to the two asymmetry terms $\sin\chi$ (index 1) and $\sin 2\chi$ (index 2). $\Delta A_i^{(a)}$ is the difference between the expected asymmetry calculated from the simulated templates and the measured $A_i^{(a)}$: 
\begin{equation}
    \Delta A_i^{(a)} = 
    \frac{\Im(g_R)_{\rm fit}}{\Im(g_R)_{0}}A_{RH,i}^{(a)} + 
    \frac{\Im(g_P g_T^*)_{\rm fit}}{\Im(g_P g_T^*)_{0}}A_{PT,i}^{(a)} 
    - A_i^{(a)}. 
    \label{eq:template_diff}
\end{equation}
For each bin $i$, the $2\times 2$ matrix $\Sigma^{(ab)}_i$ is an estimate of the covariance between the $\sin\chi$ and $\sin 2\chi$-weighted asymmetries in the bin $i$: \begin{equation}
    \Sigma_{i} = \left(\frac{N_{\rm bins}}{N_{\rm signal}}\right)^2\left(\begin{array}{cc}
        \sum\sin^2\chi_n & \sum\sin\chi_n \sin 2\chi_n \\
        \sum\sin\chi_n \sin 2\chi_n  & \sum\sin^2 2\chi_n
    \end{array}\right), 
\end{equation}
where the summation is performed over all the events in the bin $i$: $\sum\equiv\sum_{n=1}^{N_i}$. In Eq.~\ref{eq:template_diff}, $A_{RH,i}^{(1)}$ and $A_{PT,i}^{(1)}$ are the binned $\sin\chi$-weighted templates for the asymmetries due to right-handed current (with the coupling $\Im(g_R)_0$) and pseudoscalar-tensor interference (with the coupling $\Im(g_P g_T^*)_0$) obtained from simulation. Similarly, the $A_{RH,i}^{(2)}$ and $A_{PT,i}^{(2)}$ are the asymmetry templates with $\sin 2\chi$ weight. $A_i^{(1)}$ and $A_i^{(2)}$ are the binned asymmetries observed in data. 

\begin{figure}
    \centering
    \includegraphics[width=0.48\textwidth]{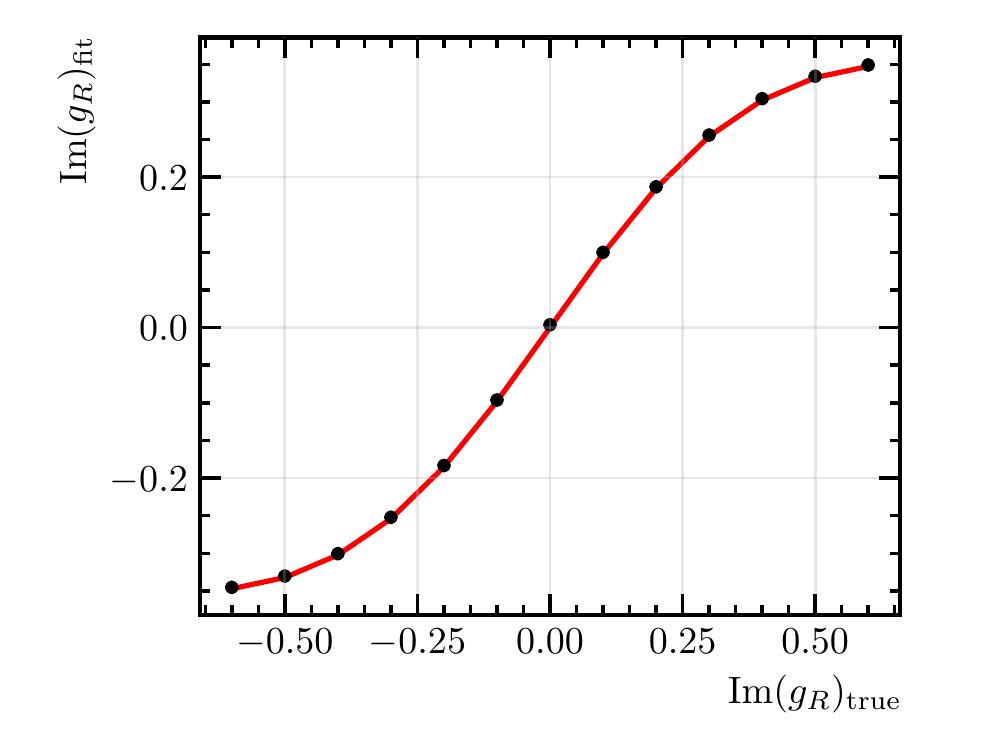}
    \put(-159, 131){(a)}
    \includegraphics[width=0.48\textwidth]{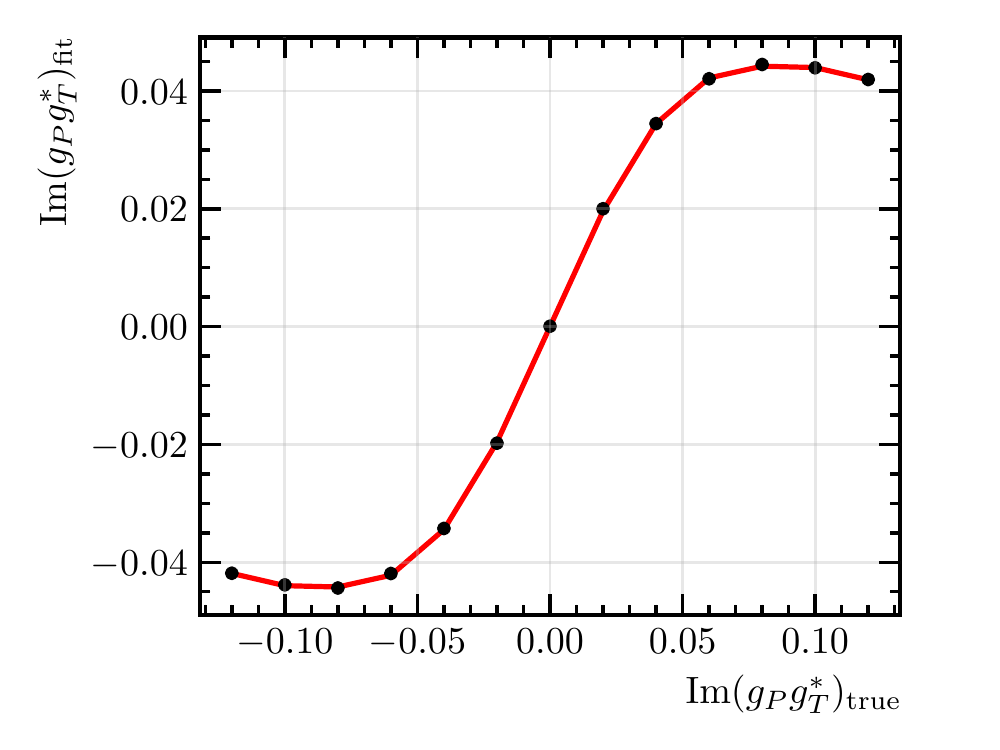}
    \put(-159, 131){(b)}
    \caption{Dependence of the NP coupling obtained from the binned template fits on the coupling used in the simulation (``true'') for (a) right-handed current contribution and (b) interference of pseudoscalar and tensor currents. }
    \label{fig:linearity}
\end{figure}

Note that in Eq.~\ref{eq:chi2_fit}, the term $A_{PT}^{(2)}$ is kept that is missing in the truth-level distribution (\ref{eq:p_odd_expr}). That is because in general the non-uniform acceptance or topological reconstruction of the $\chi$ angle could introduce higher harmonics to the $P$-odd NP terms that are proportional to $\sin\chi$. 
Equations~\ref{eq:p_odd_expr} and their binned version (\ref{eq:template_diff}) are linear in the NP couplings, which assumes that the NP admixture to the SM part of the density is small. For sufficiently large NP contribution, the terms quadratic in $g_R$ and $g_Pg_T^*$ will become non-negligible, and the linearity will no longer hold. The interval of values of the NP couplings where the linearity assumption is still valid is investigated with simulated samples where NP couplings are scanned in a broad range, and the linear fits using Eq.~\ref{eq:chi2_fit} are performed. The dependence of the fitted NP couplings on the input ones is shown in Fig.~\ref{fig:linearity}. One can see that saturation of the linear regime occurs at around $|\Im(g_R)|>0.2$ and $|\Im(g_Pg_T^*)|>0.03$.\footnote{In the case of pseudoscalar-tensor interference, the linear regime requires that both pseudoscalar and tensor contributions to the decay amplitude are small with respect to the SM part. The results shown here correspond to the case $g_P=1$ and $g_T$ scanned in the range from $-0.12i$ to $+0.12i$. For other combinations, the linear range will be different, but what is important is that if the linearity holds, then the asymmetry only depends on the combination $\Im(g_Pg_T^*)$ and not on the individual values of the two couplings. } In order not to bias the NP couplings in the linear fit, all the subsequent studies are performed with the asymmetry templates $A^{(a)}_{RH,i}$ and $A^{(a)}_{PT,i}$ obtained from the simulated samples in the linear region: $\Im(g_R)_0 = 0.1$ and $\Im(g_Pg_T^*)_0=0.02$. 

Instead of dealing with weighted histograms as in Eq.~\ref{eq:binned_asym}, one could cancel out the $P$-even part by splitting the sample in $\chi<0$ and $\chi>0$ parts and taking the difference of yields in bins. This is equivalent to using the weights equal to ${\rm sign}(\sin \chi)$ rather than $\sin\chi$, and is similar to dealing with triple-product asymmetries that is a commonly used technique in fully reconstructed beauty and charm decays~\cite{Gronau:2011cf}. In order to access the $\sin 2\chi$ term, one needs to count the number of events in four quadrants of $\chi$ (\ie use the weight ${\rm sign}(\sin 2\chi)$). This approach is considered in further study of statistical sensitivity. 

\section{Statistical sensitivity}

\label{sec:stat}

A simulation study has been performed in order to estimate the statistical sensitivity of the proposed technique. Each simulated sample used in the study contains $5\times 10^6$ events satisfying the selection criteria listed in Section~\ref{sec:simulation}. This is approximately 15 times larger than the yield of $\Bzb\to \Dstarp\mun\neumb$ decays observed by LHCb in the data sample corresponding to 3\invfb integrated luminosity of LHC Run~1~\cite{LHCb:2015gmp} and roughly corresponds to the expected yield of these decays in the 50\invfb sample after Run~4 assuming the same trigger and reconstruction efficiency as in the published analysis. The asymmetry templates are also obtained from the simulated samples with approximately 4 times larger data samples using $\Im(g_R)=0.1$ (for the RH template) and $\Im(g_P g_T^*)=0.02$ (for the PT template), such that the uncertainties of the templates are smaller than those of the fitted samples. 

The statistical uncertainty for the binned fit to the SM-like sample (\ie, with all NP couplings set to zero) is shown in Table~\ref{tab:stat} for the default $8\times 8$ binning. The precision from the distributions using ``true'' angles is compared with those obtained from the three solutions of the $p_B$ momentum discussed in Section~\ref{sec:reconstruction}. In all cases, the same solution is used for the fitted data and for the construction of asymmetry templates. The precision for the $\Im(g_R)$ and $\Im(g_P g_T^*)$ couplings from the combined fit using the $\chi^2_{\rm corr}$ function (\ref{eq:chi2_fit}) are shown in the columns ``RH Combined'' and ``PT combined'', respectively. In addition, the values of $\Im(g_R)$ obtained from the $\sin\chi$ and $\sin 2\chi$ asymmetry terms separately are given in columns ``RH $\sin\chi$'' and ``RH $\sin 2\chi$''. Finally, the $\Im(g_P g_T^*)$ coupling is mostly constrained by the $\sin\chi$ asymmetry: the precision using this term only is given in the column titled ``PT $\sin\chi$''. $\sin2\chi$ term alone does not provide meaningful constraint of the PT term and thus the corresponding precision is not presented separately. 

\begin{table}
  \caption{Statistical precision for the imaginary parts of the right-handed coupling $g_R$ (RH) and the combination of pseudoscalar and tensor couplings $g_P g_T^*$ (PT) with the template fits using different solutions for the $B$ momentum estimation. }
  \label{tab:stat}
  \begin{center}
  \begin{tabular}{lrrrrr}
\toprule
      Solution &  RH $\sin\chi$ &  RH $\sin 2\chi$ &  RH Combined &  PT $\sin\chi$ &  PT Combined \\
\midrule
          True &        0.00252 &          0.00190 &      0.00143 &        0.00021 &      0.00020 \\
Solution '$-$' &        0.00372 &          0.00497 &      0.00219 &        0.00029 &      0.00028 \\
Solution '$+$' &        0.00569 &          0.01017 &      0.00375 &        0.00045 &      0.00045 \\
Solution 'avg' &        0.00464 &          0.00978 &      0.00308 &        0.00036 &      0.00036 \\
\bottomrule
\end{tabular}

  \end{center}
\end{table}

\begin{figure}
    \centering
    \includegraphics[width=0.48\textwidth]{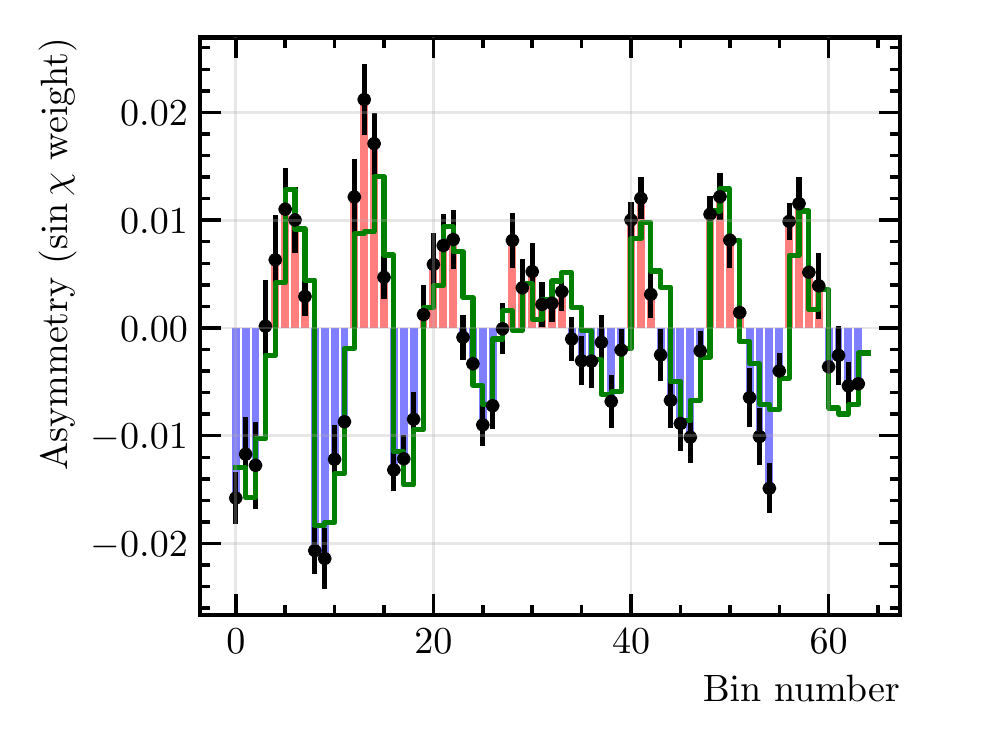}
    \put(-43, 41){(a)}
    \includegraphics[width=0.48\textwidth]{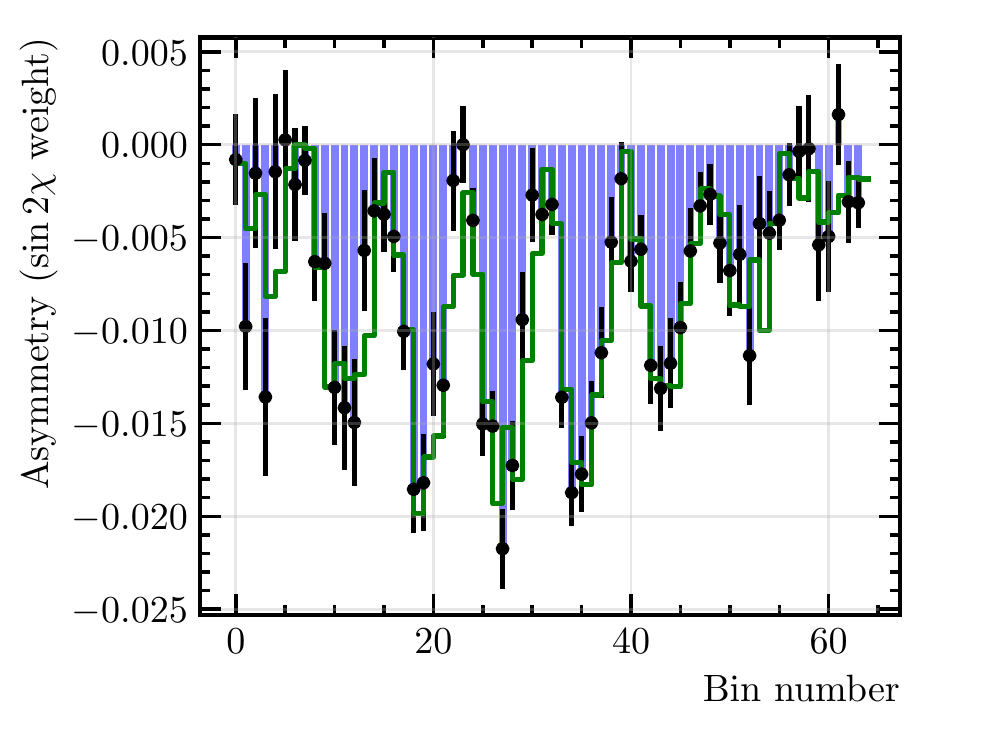}
    \put(-43, 41){(b)}
    \caption{Flattened $8\times 8$ binned (a) $\sin\chi$ and (b) $\sin 2\chi$ asymmetries for the amplitude with 
    right-handed current ($\Im(g_R)=0.1$) using reconstructed decay parameters (solution ``$-$''). 
    Points with error bars represent the asymmetry obtained from the simulation, and the solid-line histogram is the result of template fit. }
    \label{fig:comb_fit_rh}
\end{figure}

\begin{figure}
    \centering
    \includegraphics[width=0.48\textwidth]{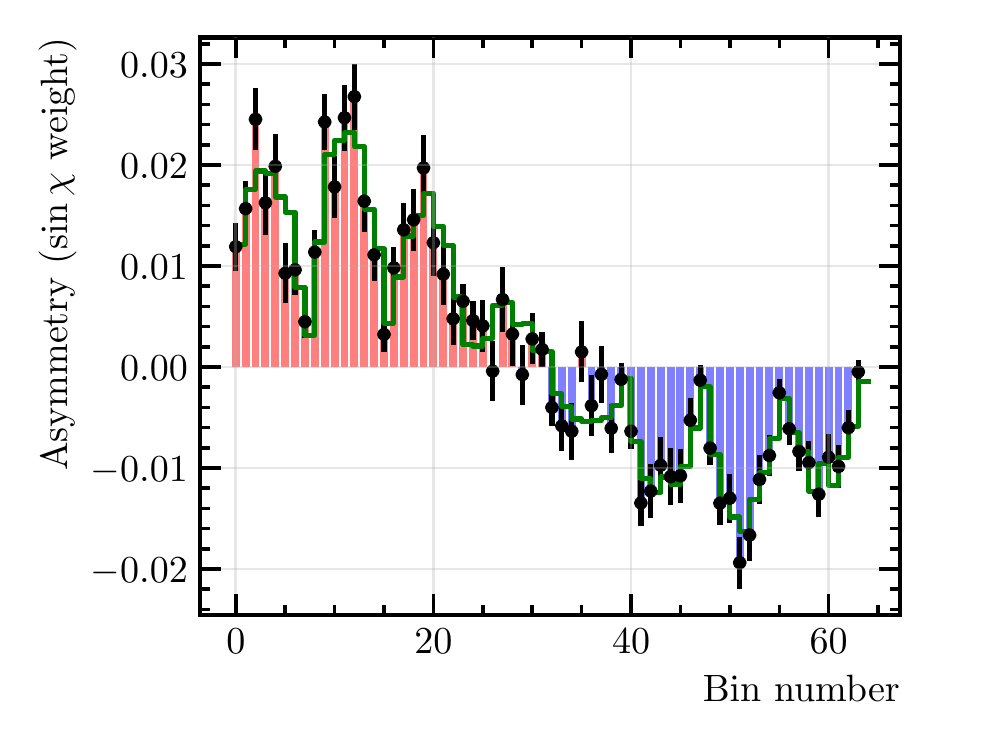}
    \put(-43, 131){(a)}
    \includegraphics[width=0.48\textwidth]{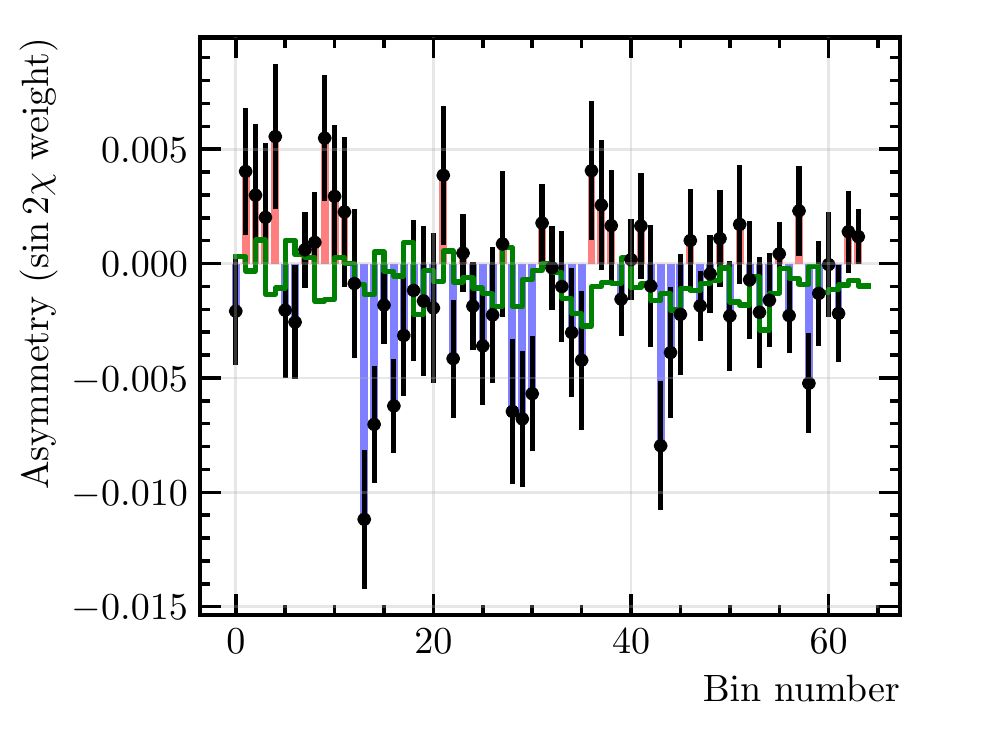}
    \put(-43, 131){(b)}
    \caption{Flattened $8\times 8$ binned (a) $\sin\chi$ and (b) $\sin 2\chi$ asymmetries for the amplitude with 
    the interference of pseudoscalar and tensor currents 
    ($\Im(g_P g_T^*)=0.02$) using reconstructed decay parameters (solution ``$-$''). 
    Points with error bars represent the asymmetry obtained from the simulation, and the solid-line histogram is the result of template fit. }
    \label{fig:comb_fit_pt}
\end{figure}

Clearly, the best precision is offered by the solution ``$-$''. Therefore, in all the subsequent studies solution ``$-$'' is used. The binned $\sin\chi$ and $\sin 2\chi$ asymmetries for this solution are shown in Fig.~\ref{fig:comb_fit_rh} for the RH model with $\Im(g_R)=0.1$ and in Fig.~\ref{fig:comb_fit_pt} for the PT model with $\Im(g_P g_T^*)=0.02$. Instead of the 2D binned $8\times 8$ distributions as in Fig.~\ref{fig:asym_rh_pt}, the flattened distributions with 64 bins are shown to visualise the uncertainties of the asymmetry in bins. The bins are counted from negative to positive values of $\cos\theta_{\ell}$, and then from negative to positive values of $\cos\theta_D$. The fit result using the combined template fit with the $\chi^2_{\rm corr}$ function is also shown.

Statistical precision for different numbers of bins in the template is shown in Table~\ref{tab:binning}. One can see that increasing the number of bins from $4\times 4$ to $16\times 16$ does not result in a significant change in the precision of the binned fit. While even the rough binning is sufficient to resolve the two NP couplings, using finer binning may have an advantage since it provides more degrees of freedom to control the systematic effects, which, as shown in Sec.~\ref{sec:systematics}, produce an asymmetry pattern that is different from the one generated by NP components. 

\begin{table}
  \caption{Statistical precision for the imaginary parts of the right-handed coupling $g_R$ (RH) and the combination of pseudoscalar and tensor couplings $g_P g_T^*$ (PT) 
  for the weighted asymmetry fits with different binning. }
  \label{tab:binning}
  \begin{center}
  \begin{tabular}{lrr}
\toprule
      Binning &  RH Combined &  PT Combined \\
\midrule
  $4\times 4$ &      0.00236 &      0.00029 \\
  $6\times 6$ &      0.00224 &      0.00028 \\
  $8\times 8$ &      0.00219 &      0.00028 \\
$12\times 12$ &      0.00213 &      0.00027 \\
$16\times 16$ &      0.00208 &      0.00027 \\
\bottomrule
\end{tabular}

  \end{center}
\end{table}

Finally, Table~\ref{tab:split} shows the comparison of the statistical precision for the weighted asymmetry with the approach when the sample is split according to the sign of $\sin\chi$ ($\sin 2\chi$) terms as mentioned at the end of Section~\ref{sec:binned_fit}. The precision for the  approach with the split samples is marginally worse compared to the weighted one. 

\begin{table}
  \caption{Comparison of the statistical precision for the imaginary parts of the right-handed coupling $g_R$ (RH) and the combination of pseudoscalar and tensor couplings $g_P g_T^*$ (PT) 
  for the weighted asymmetry fits and the fits of the split sample asymmetry. }
  \label{tab:split}
  \begin{center}
  \begin{tabular}{lrr}
\toprule
                            Asymmetries &  RH Combined &  PT Combined \\
\midrule
           Weighted ($\sin \chi/2\chi$) &      0.00219 &      0.00028 \\
Split sample (${\rm sign}(\sin \chi/2\chi)$) &      0.00238 &      0.00030 \\
\bottomrule
\end{tabular}

  \end{center}
\end{table}

In conclusion, the statistical precision of the $CP$ violating observables is estimated to be of the order 0.2\% for the $\Im(g_R)$ coupling and 0.03\% for $\Im(g_P g_T^*)$ in the upgraded LHCb data sample corresponding to 50\invfb luminosity and assuming the similar detector performance as in Run 1 analysis~\cite{LHCb:2015gmp}. The currently available data sample of 9\invfb should thus be sufficient to measure the contribution of right-handed current with the statistical precision below 1\% and the combination of pseudoscalar and tensor couplings $\Im(g_P g_T^*)$ below 0.1\%. 

\section{Systematic effects}
\label{sec:systematics}

The systematic effects that can affect the $CP$-violating observables can be split into two groups: the parity-even and parity-odd ones. If the true distribution is SM-like, the former cannot produce fake parity-odd terms in the angular distribution. However, they affect the interpretation of the visible asymmetry in terms of theory parameters. The uncertainties in the background composition, formfactor parametrisation, reconstruction, detection, and selection efficiency belong to this group of effects. For instance, the formfactor uncertainty enters relatively to the magnitude of the $CP$-odd contribution and should be negligible if $CP$ violation in data is consistent with zero. Similarly, the uncertainty due to the purity of the data sample (the fraction of signal events) enters the normalisation term $N_{\rm bins}/N_{\rm signal}$ in the asymmetry (\ref{eq:binned_asym}) and is again proportional to the visible $CP$ asymmetry. 

The second group of systematic effects, parity-odd ones, are the most dangerous since they can produce the fake parity-odd (chiral) terms in the SM-like angular distribution. Such effects are considered in this section in more detail. 

\subsection{Backgrounds}

The decay mode $\Bzb\to\Dstarp\mun\neumb$ can be reconstructed in a clean way in the $pp$ collisions by combining the $\Dstarp$ and $\mun$ candidates~\cite{LHCb:2015gmp}. Due to the excellent charged particle identification capabilities and momentum resolution of LHCb the backgrounds with misidentified particles and combinatorial \Dstarp are low and can be subtracted in a data-driven way.\footnote{One example of the backgrounds that cannot be suppressed to a negligible level is the $\B\to \Dstarm\pip X$ decays with $\pip\to \mup\neum$ decay in flight. However, the contribution of this background can be evaluated in a data-driven way by selecting the candidates where muon tracks are required to pass the pion identification requirements.} Therefore, only the backgrounds with genuine \Dstarp and \mun particles are considered in the following. The most significant of them are presented in Table~\ref{tab:background_fractions}. The estimates of the branching fractions to the $\Dstarp\mun X$ final state are presented as products of branching fractions from the PDG~\cite{PDG}. The fraction of each background with respect to the signal mode is also presented assuming that the reconstruction efficiency is the same as for the signal mode. In a real analysis, the contributions of various partially reconstructed $\B\to \Dstarp\mun X$ backgrounds can be suppressed by applying isolation requirements, such that the background fractions in Table~\ref{tab:background_fractions} present the worst-case scenario. 

\begin{table}
    \caption{The most significant expected backgrounds for the $\Bzb\to\Dstarp\mun\neumb$ decay mode, and estimates of their branching ratios (\BR)~\cite{PDG} and yield fractions with respect to signal.
    The \BR estimates for the modes with $\D^{**+}\to \Dstarp\piz$ include the $1/2$ isospin factor wrt. $\D^{**0}\to \Dstarp\pim$ probability. The $\Bsb$ decay mode includes a 25\% factor due to the ratio of $\Bsb$ and $\B$ fragmentation fractions. The $\BR(\B\to \D^{**}\taum\neutb)$ estimates are based on the respective muon modes with the SM $R(D^{**})$ factor of $\sim 0.2$~\cite{Bernlochner:2017jxt}. The contribution of $\B\to\Dstarm\D^{(*)}K$ decays is the sum over several modes whose \BR's are measured in Ref.~\cite{BaBar:2010tqo}. }
    \label{tab:background_fractions}
    \begin{center}
    \scalebox{0.92}{
    \begin{tabular}{lrr}
         \toprule
         Decay mode & \BR estimate & Fraction \\
         \midrule
         $\Bzb\to\Dstarp\mun\neumb$ (signal) & 
           $(4.97\pm 0.12)\%$ & 1 \\
         $\Bm\to(\D^{**0}\to\Dstarp\pim)\mun\neumb$ & 
           $(6.0\pm 0.4)\times 10^{-3}$ & 0.12 \\
         $\Bzb\to(\D^{**+}\to\Dstarp\piz)\mun\neumb$ & 
           $1/2\times (6.0\pm 0.4)\times 10^{-3}$ & 0.06 \\
         $\Bzb\to\Dstarp\taum\neutb$ & 
           $(1.58\pm 0.09)\%\times (17.39\pm 0.04)\%$ & 0.055 \\
         $\Bsb\to(\D_{s1}^{+}\to\Dstarp K^0)\mun\neumb$ & 
           $25\%\times 2\times (2.7\pm 0.7)\times 10^{-3}$ & 0.027 \\

         $\Bm\to(\D^{**0}\to\Dstarp\pim)\taum\neutb$ & 
           $0.2\times (6.0\pm 0.4)\times 10^{-3}\times (17.39\pm 0.04)\%$ & $0.004$\\
         $\Bzb\to(\D^{**+}\to\Dstarp\piz)\taum\neutb$ & 
           $1/2\times 0.2\times (6.0\pm 0.4)\times 10^{-3}\times (17.39\pm 0.04)\%$ & $0.002$\\

         \midrule

         $\Bzb\to \Dstarp\D_s^{*-}$, $\Dsm\to \mun\neumb X$ & 
           $(1.77\pm 0.14)\% \times 0.94 \times (6.33\pm 0.15)\%$ & 0.021 \\
         $\Bzb\to \Dstarp\Dsm$, $\Dsm\to \mun\neumb X$  & 
           $(8.0\pm 1.1)\times 10^{-3} \times (6.33\pm 0.15)\%$ & 0.010 \\
           
         $\B\to \Dstarp\D^{(*)}K$, $\D\to \mun\neumb X$ & 
           $4\times 10^{-2} \times (6.8\pm 0.6)\%$ & $0.05$ \\

         $\B\to \Dstarp\D_s^{(*)-}\pi$, $\Dsm\to \mun\neumb X$ & 
           $2/3\times (2.7\pm 1.1)\% \times 10^{-2} \times (6.33\pm 0.15)\%$ & $0.02$ \\

         \bottomrule
    \end{tabular}
    }
    \end{center}
\end{table}

The backgrounds in Table~\ref{tab:background_fractions} can be split into two groups: the semileptonic $\B$ decays with either $\mun\neumb$ or $\taum\neutb$ combinations (in the latter case $\taum$ decays into $\mun\neumb\neut$), and double charm decays with the $\Dstarm$ and another charm hadron in the final state, followed by a semileptonic decay of the second charm hadron. Semileptonic decays do not produce $CP$ violation in the SM, so they can only affect the interpretation of the $CP$-violating experimental observables in terms of NP couplings. Some of those backgrounds, however, can produce parity-odd (but not $CP$-odd) terms. On the other hand, double charm backgrounds are, in general, $CP$-violating, although $CP$ asymmetry in the dominant ones is expected to be small. 

\begin{figure}
    \centering
    \includegraphics[width=0.48\textwidth]{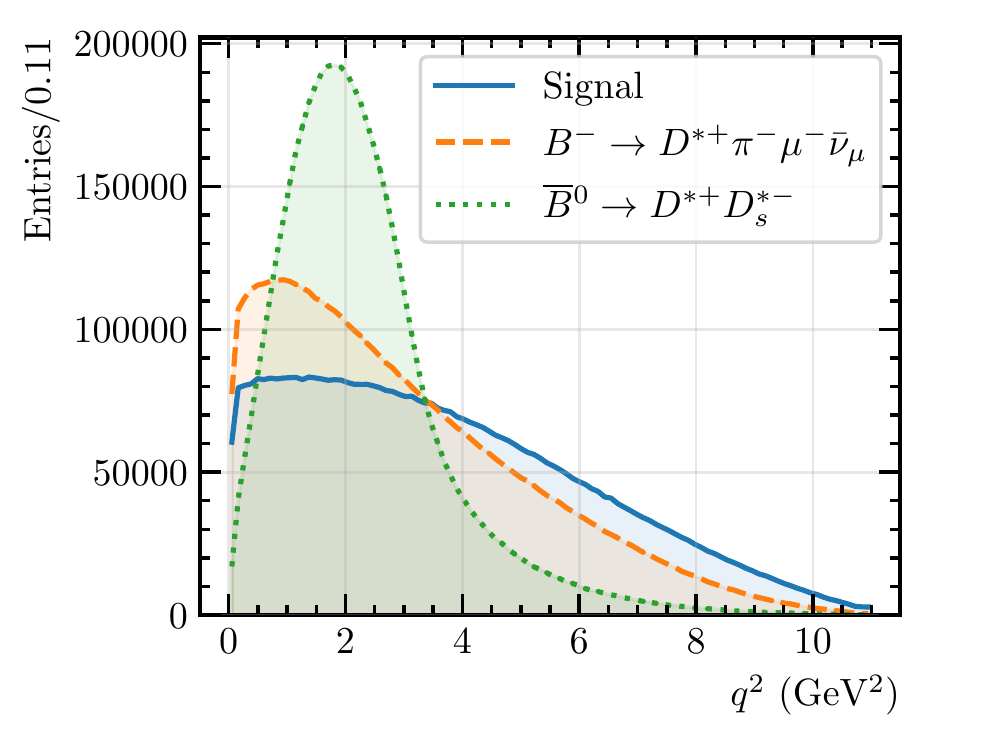}

    \caption{Reconstructed $q^2$ distributions for the signal and backgrounds. }
    \label{fig:bck_q2}
\end{figure}

The reconstructed $q^2$ distribution for the backgrounds is different from that of the signal decays. The example of $q^2$ distributions for the two backgrounds that are mentioned below, semileptonic decays with additional pion and doubly charmed decays, is shown in Fig.~\ref{fig:bck_q2}. As a result, it could be useful to perform the study of $CP$ asymmetries in at least two $q^2$ regions to control possible background contributions to the asymmetry. 

\subsubsection{Semileptonic backgrounds}
\label{sec:sl_bck}

The dominant partially reconstructed semileptonic backgrounds for the $\Bzb\to\Dstarp\mun\neumb$ mode are coming from the decays with higher $\D$ excitations ($\D^{**}$), either charged or the neutral ones, with the neutral or charged pion from the $\D^{**}\to \Dstarm\pi$ decay not reconstructed. Here $\D^{**}$ is the admixture of several states with different quantum numbers. As a consequence of that, the decay $\B\to\D^{**}\mun\neumb$ can exhibit a different mechanism of $CP$ violation, where the strong phase difference that is necessary for the $CP$ asymmetry is generated in the interference of strong decays of different $\D^{**}$ states~\cite{Aloni:2018ipm}.

To estimate the effect of the $\B\to\D^{**}\mun\neumb$ background, the $\Bm\to\Dstarp\pim\mun\neumb$ events are generated where the $\Dstarp\pim$ combination is coming from the interference of $D_1(2420)$ (with quantum numbers $J^P=1^+$) and $D_2^*(2460)$ ($J^P=2^+$) states. The decay amplitudes for the two states follow the formalism from Ref.~\cite{Bernlochner:2017jxt}. Since the strong phase difference between the $D_1(2420)$ and $D_2^*(2460)$ states is unknown, eight data samples are generated with the phase difference $\delta_D$ between them changing from $0^{\circ}$ to $315^{\circ}$ in steps of $45^{\circ}$. 
The procedure similar to the one for the simulation of the $\Bzb\to\Dstarp\mun\neumb$ signal (Sec.~\ref{sec:simulation}) is used to simulate the production of $\Bm$ mesons, calculate the momenta of the reconstructed decay products in the laboratory frame, smear and reconstruct the $\Bm$ decay vertex and then reconstruct the parameters $q^2$, $\cos\theta_D$, $\cos\theta_{\ell}$ and $\chi$ in the assumption that the decay products come from the $\Bzb\to\Dstarp\mun\neumb$ mode. Finally, the binned asymmetries are calculated and fitted with the RH and PT templates (Sec.~\ref{sec:binned_fit}). 

\begin{figure}
    \centering
    \includegraphics[width=0.48\textwidth]{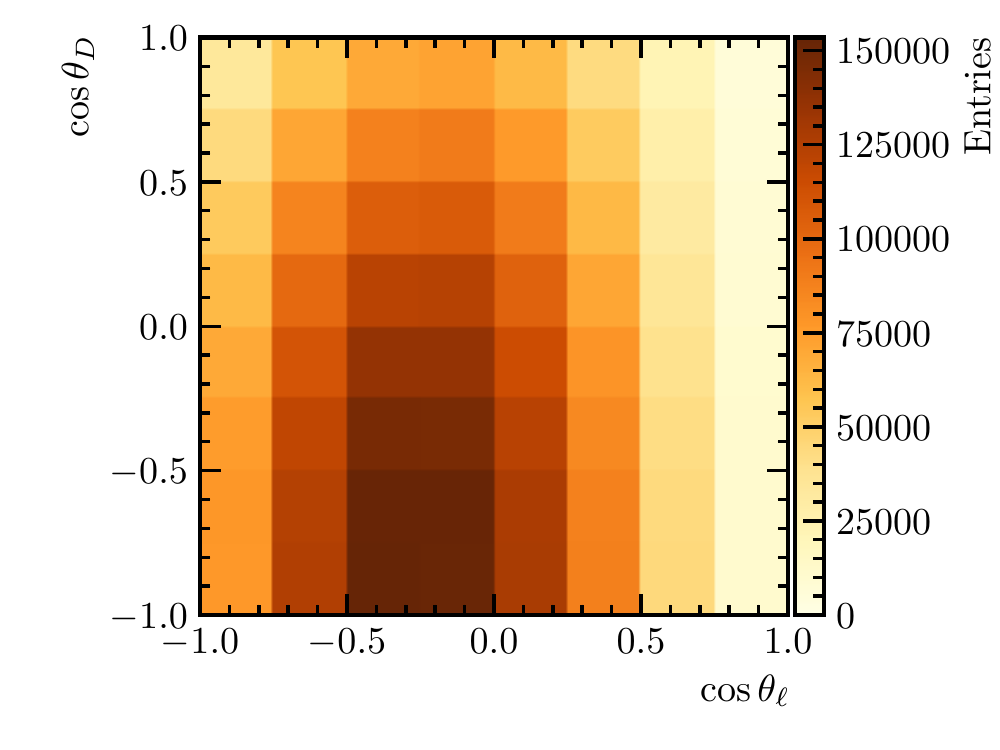}
    \put(-159, 131){\colorbox{white}{(a)}}

    \includegraphics[width=0.48\textwidth]{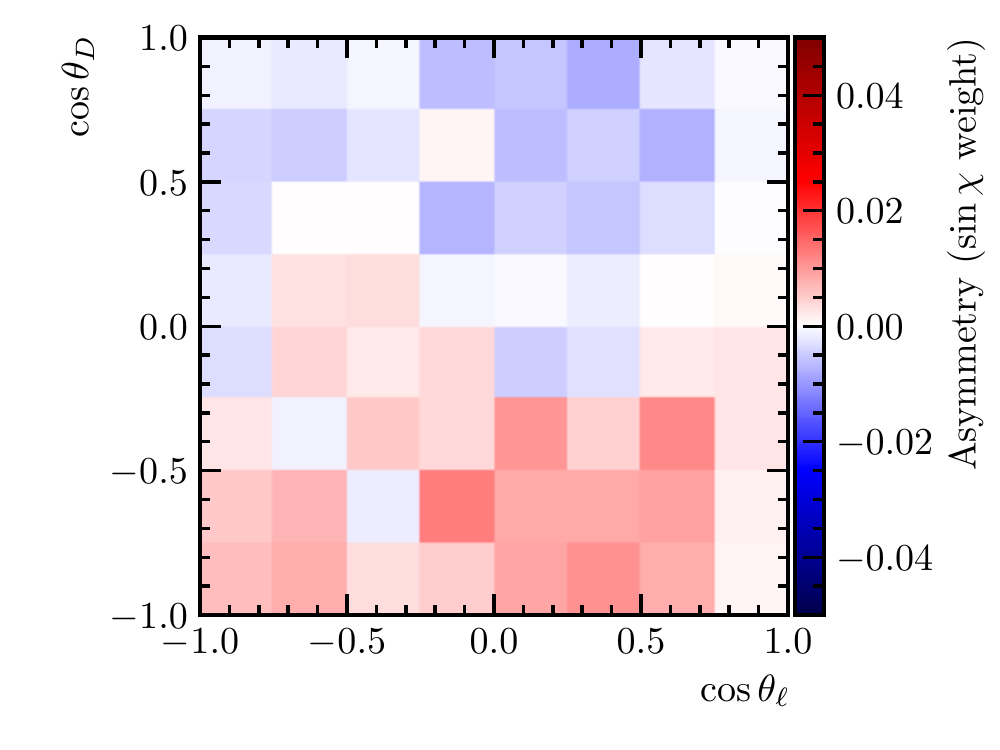}
    \put(-159, 131){\colorbox{white}{(b)}}
    \includegraphics[width=0.48\textwidth]{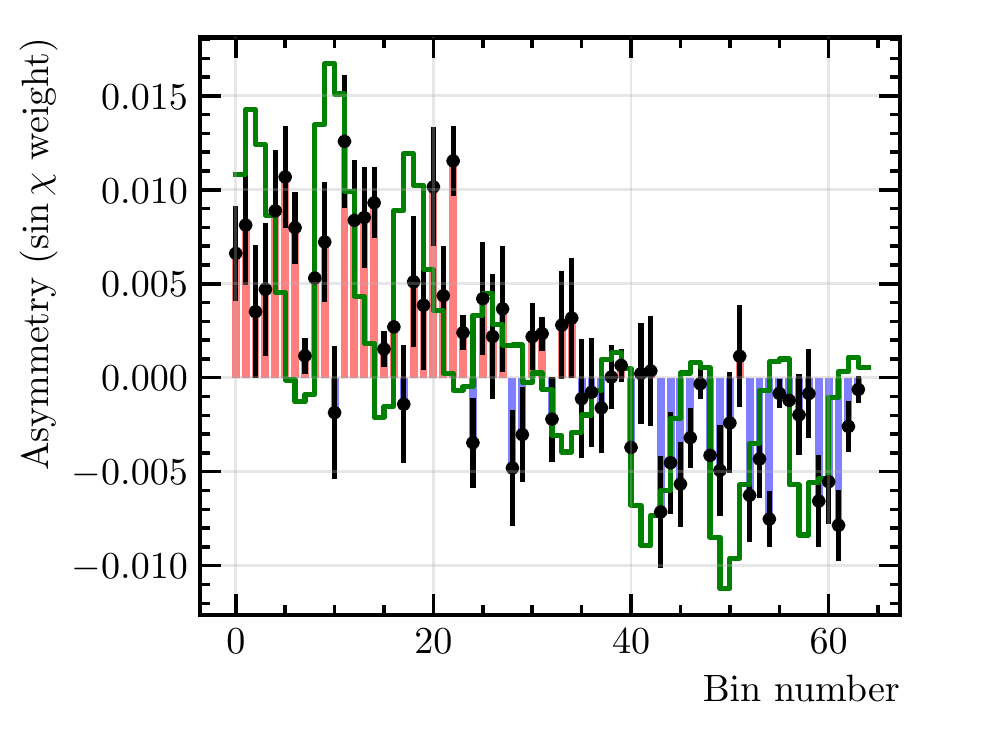}
    \put(-43, 131){\colorbox{white}{(c)}}

    \includegraphics[width=0.48\textwidth]{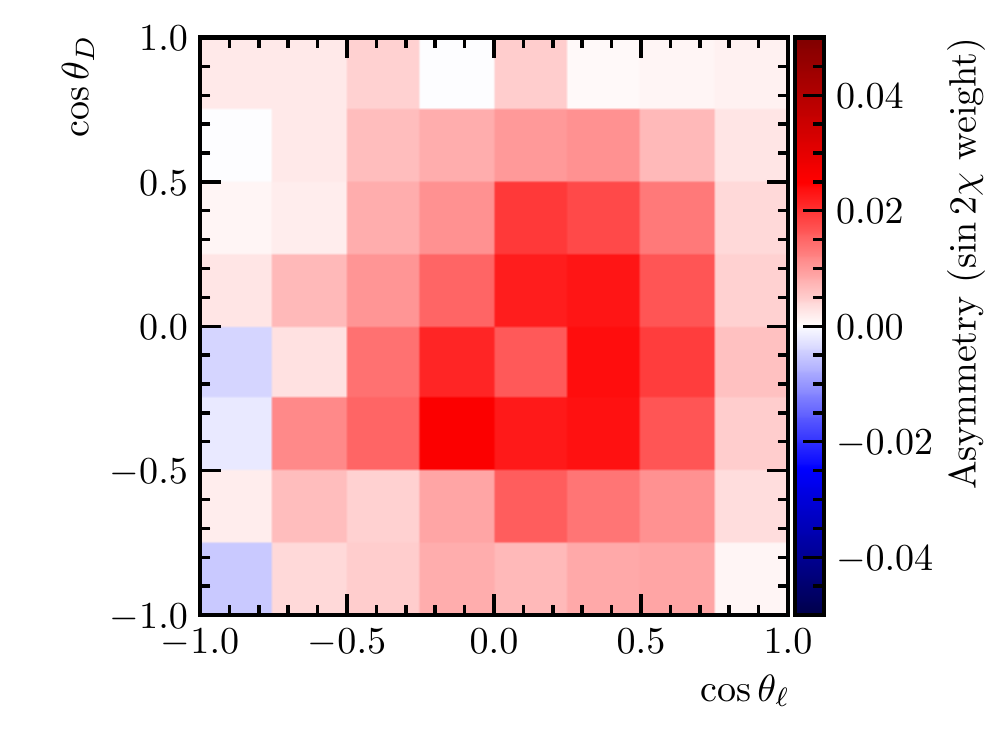}
    \put(-159, 131){\colorbox{white}{(d)}}
    \includegraphics[width=0.48\textwidth]{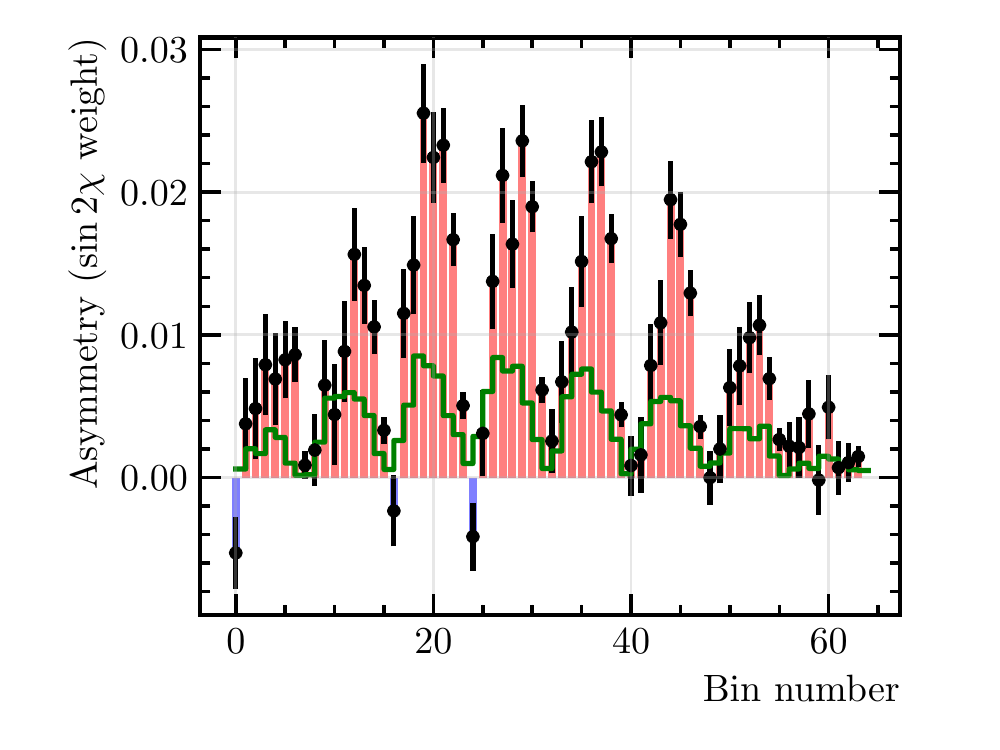}
    \put(-43, 131){\colorbox{white}{(e)}}

    \caption{(a) Binned density and binned asymmetries of (b,c) $\sin\chi$ and (d,e) $\sin 2\chi$
    terms for $\Bm\to \Dstarp\pim\mun\neumb$ background with $\delta_D=315^{\circ}$. Plots (b) and (d) show 2D binned asymmetries, (c) and (e) are the corresponding flattened asymmetries (points with error bars), and the result of the best fit using RH and PT templates (solid-line histogram). }
    \label{fig:dstst_bck}
\end{figure}

The fitted values of the RH and PT couplings as a function of the strong phase difference are shown in Table~\ref{tab:bck_bias}. The asymmetry pattern in $\cos\theta_D$ vs. $\cos\theta_{\ell}$ bins for $\delta_D=315^{\circ}$ (the value that gives close to maximal asymmetry) and the results of the template fit are shown in Fig.~\ref{fig:dstst_bck}. The pattern differs from that seen in the case of RH or PT NP terms (Fig.~\ref{fig:asym_rh_pt}) such that, although the fitted values of the NP couplings differ from zero, the fit quality is poor. The values of the asymmetry are normalised to the number of background decays; since in the experiment this background is expected to contribute at the level of 20\%, the worst-case bias if this background is not accounted for (assuming that both the $\D^{**0}$ and $\D^{**+}$ have the same strong phase difference that gives the maximum effect) is of the order $\Delta \Im(g_R)\simeq 0.01$ and $\Delta \Im(g_P g_T^*)\simeq 0.001$. 

\begin{table}
  \caption{Fitted values of the NP couplings from the binned asymmetry fit for the 
           background samples}
  \label{tab:bck_bias}
  \begin{center}
  \begin{tabular}{lrr}
\toprule
{Background sample} & {$\Delta\Im(g_R)$} & {$\Delta\Im(g_P g_T^*)$} \\
\midrule
$B^-\to D^{\ast +}\pi^{-}\mu^{-}\bar{\nu}_{\mu}$, $\phi=0^{\circ}$ & $-0.0223 \pm 0.0024$ & $0.00142 \pm 0.00027$ \\
$B^-\to D^{\ast +}\pi^{-}\mu^{-}\bar{\nu}_{\mu}$, $\phi=45^{\circ}$ & $0.0079 \pm 0.0024$ & $-0.00092 \pm 0.00027$ \\
$B^-\to D^{\ast +}\pi^{-}\mu^{-}\bar{\nu}_{\mu}$, $\phi=90^{\circ}$ & $0.0334 \pm 0.0024$ & $-0.00363 \pm 0.00027$ \\
$B^-\to D^{\ast +}\pi^{-}\mu^{-}\bar{\nu}_{\mu}$, $\phi=135^{\circ}$ & $0.0404 \pm 0.0024$ & $-0.00331 \pm 0.00027$ \\
$B^-\to D^{\ast +}\pi^{-}\mu^{-}\bar{\nu}_{\mu}$, $\phi=180^{\circ}$ & $0.0258 \pm 0.0024$ & $-0.00166 \pm 0.00026$ \\
$B^-\to D^{\ast +}\pi^{-}\mu^{-}\bar{\nu}_{\mu}$, $\phi=225^{\circ}$ & $-0.0075 \pm 0.0024$ & $0.00101 \pm 0.00026$ \\
$B^-\to D^{\ast +}\pi^{-}\mu^{-}\bar{\nu}_{\mu}$, $\phi=270^{\circ}$ & $-0.0295 \pm 0.0024$ & $0.00364 \pm 0.00026$ \\
$B^-\to D^{\ast +}\pi^{-}\mu^{-}\bar{\nu}_{\mu}$, $\phi=315^{\circ}$ & $-0.0443 \pm 0.0024$ & $0.00344 \pm 0.00027$ \\
$\overline{B}{}^0\to D^{\ast +} D^{\ast -}_s$ & $-0.0001 \pm 0.0004$ & $\left(2 \pm 6\right) \times 10^{-5}$ \\
\bottomrule
\end{tabular}

  \end{center}
\end{table}

The contribution of the $\Bm\to\Dstarp\pim\mun\neumb$ background can be included in the asymmetry fit: the shape of its template is independent of the strong phase difference $\delta_D$, with only the magnitude being the function of $\delta_D$. Only the interference between $D_1(2420)$ and $D_2^*(2460)$ was considered; the other excited $\D$ states can contribute to the asymmetry to a less extent. 

It is important to remind that the discussion above is concerning parity, not $CP$ asymmetry. $CP$ asymmetry is non-zero for this background only in NP scenarios (although these scenarios can be different from those that introduce nonzero $CP$ violation in the signal decay, \eg, pure scalar or tensor couplings can also introduce $CP$ asymmetry, see Ref.~\cite{Aloni:2018ipm}). In a real analysis, it might make sense to consider independent contributions for this background for different $\B$ flavours: that will provide an additional NP observable. 

Other semileptonic backgrounds are expected to have fractions several times smaller and, even if parity violation appears to be significant for them, will probably introduce biases at a level smaller than statistical precision. 

\subsubsection{Double charm backgrounds}

The most significant double charm backgrounds are $\Bzb\to\Dstarp\Dsm$ and $\Bzb\to\Dstarp\D^{*-}_s$, $\D^{*-}_s\to \Dsm\gamma$ with semileptonic decays of the $\Dsm$ meson~\cite{LHCb:2021sqa}. The former is the decay with three spin-zero particles in the final state and thus cannot exhibit parity violation. The latter, however, has four particles in the final state and can be $P$-violating. 

\begin{figure}
    \centering
    \includegraphics[width=0.48\textwidth]{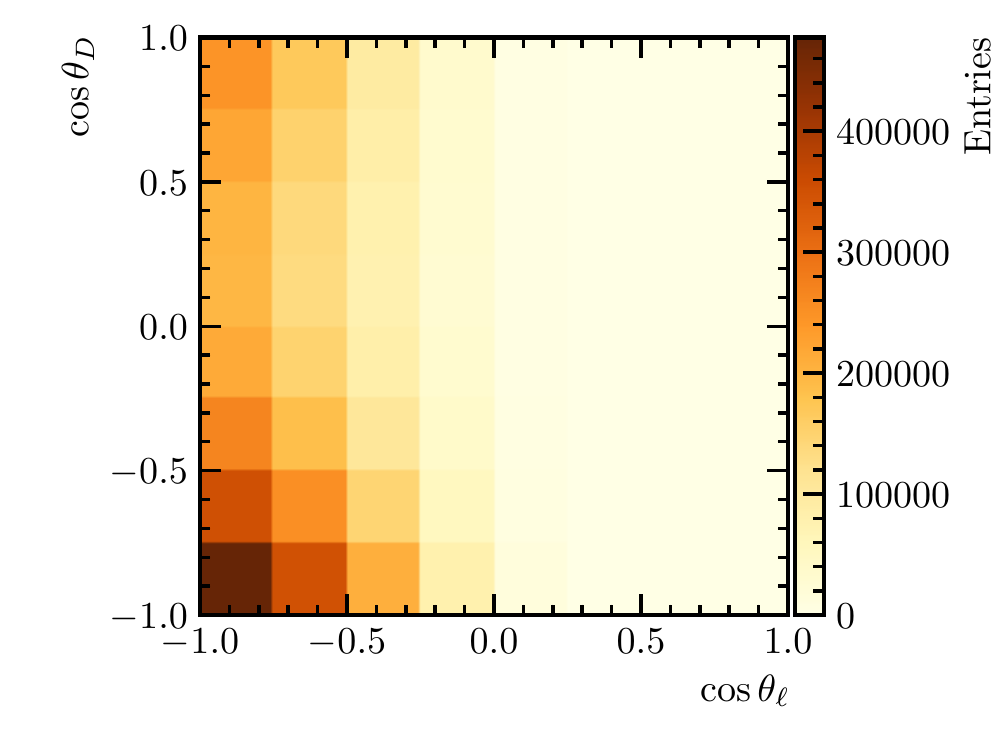}
    \put(-73, 41){\colorbox{white}{(a)}}

    \includegraphics[width=0.48\textwidth]{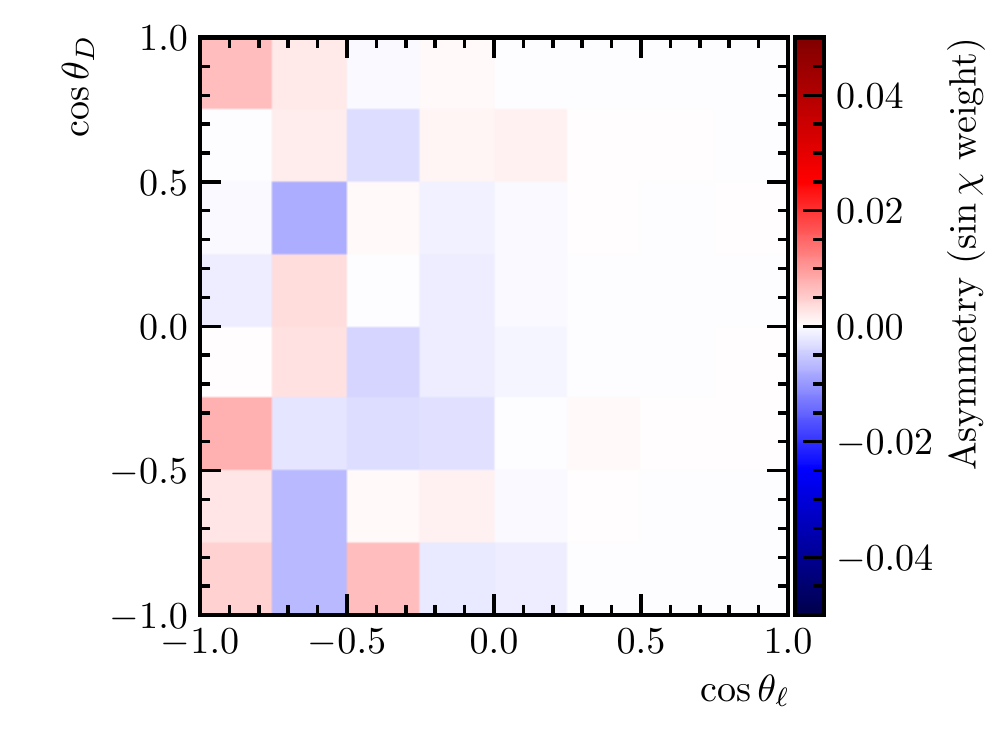}
    \put(-73, 41){\colorbox{white}{(b)}}
    \includegraphics[width=0.48\textwidth]{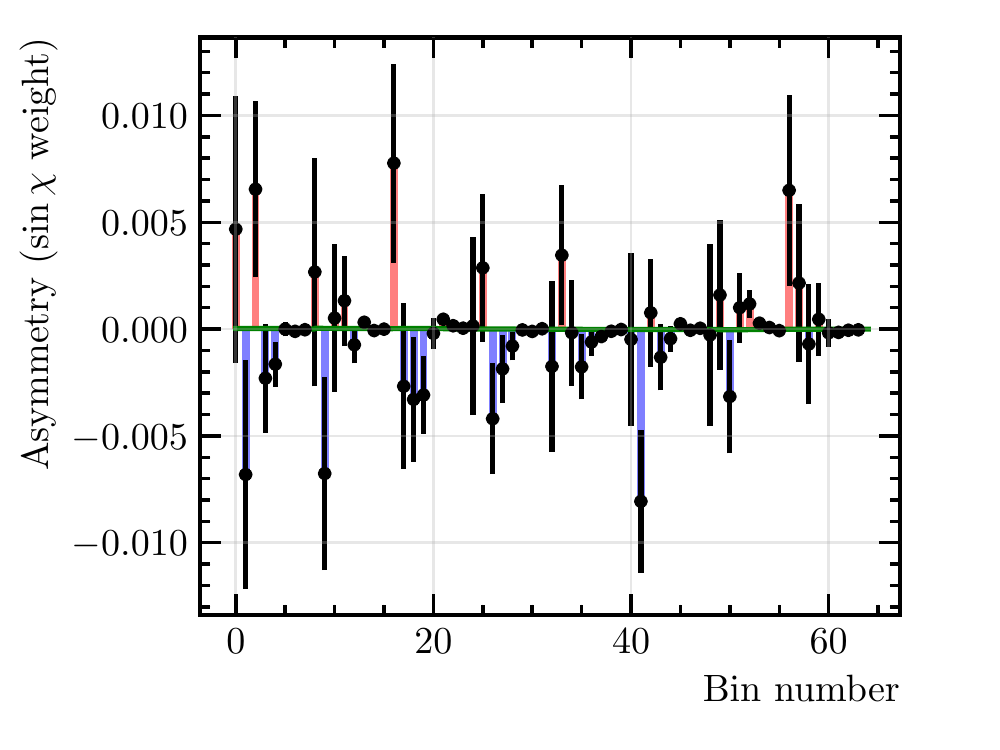}
    \put(-43, 41){(c)}

    \includegraphics[width=0.48\textwidth]{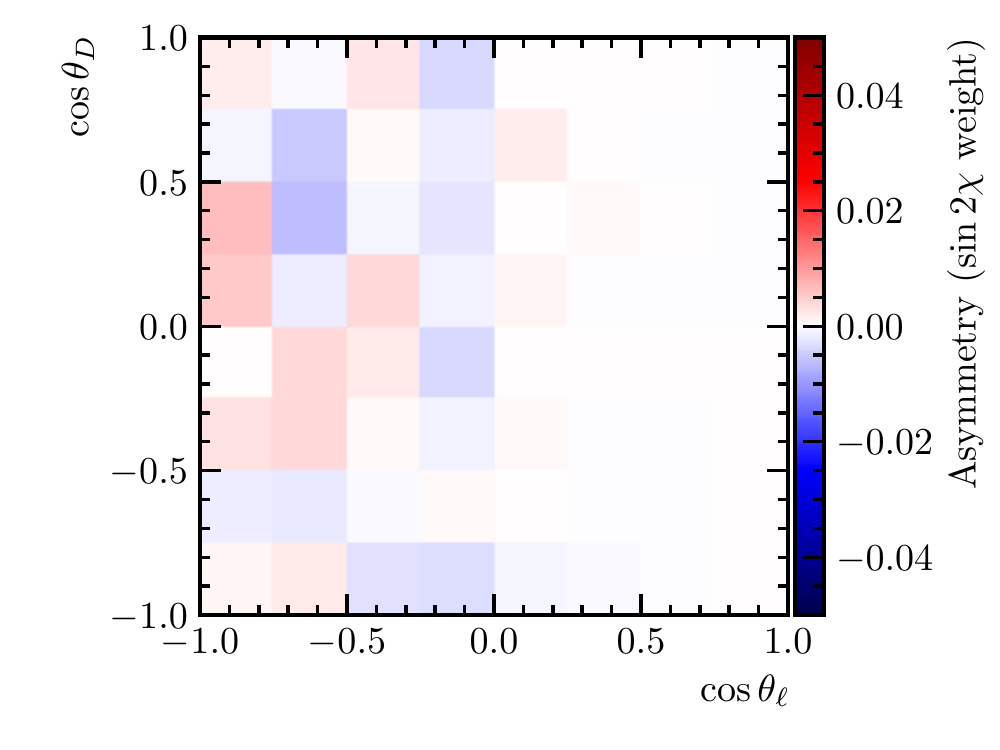}
    \put(-73, 41){\colorbox{white}{(d)}}
    \includegraphics[width=0.48\textwidth]{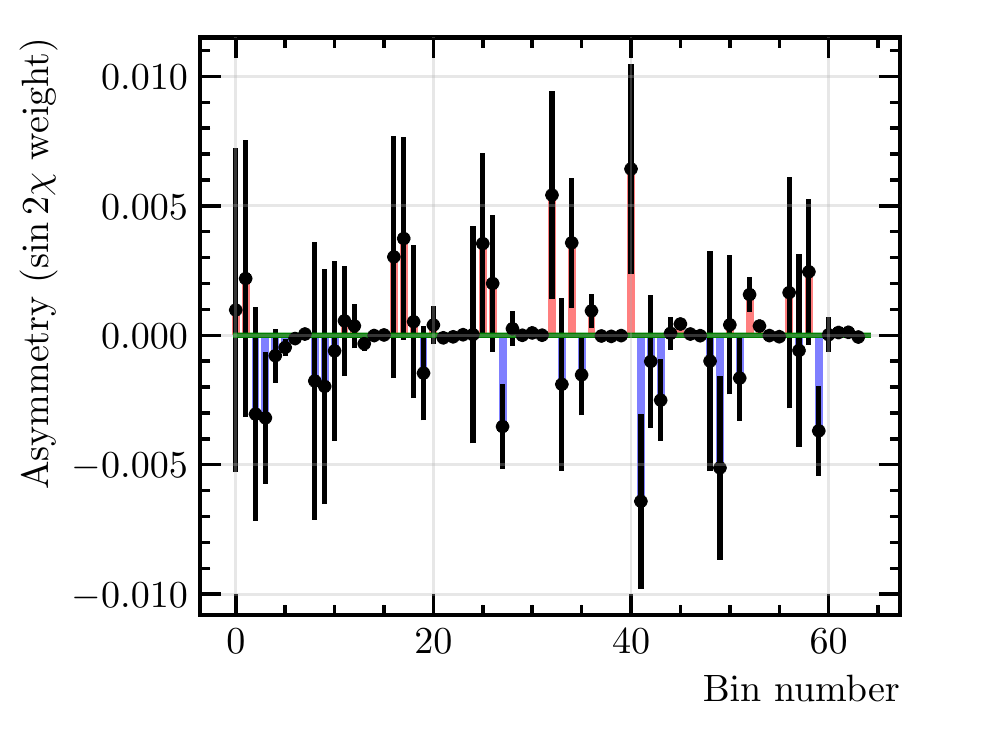}
    \put(-43, 41){(e)}

    \caption{(a) Binned density and binned asymmetries of (b,c) $\sin\chi$ and (d,e) $\sin 2\chi$
    terms for $\Bzb\to \Dstarm\D_s^{*+}$ background. Plots (b) and (d) show 2D binned asymmetries, (c) and (e) are the corresponding flattened asymmetries (points with error bars), and the result of the best fit using RH and PT templates (solid-line histogram). }
    \label{fig:dstdsst_bck}
\end{figure}

The amplitude structure of the $\Bzb\to\Dstarp\D^{*-}_s$, $\D^{*-}_s\to \Dsm\gamma$ decay has been measured by LHCb~\cite{LHCb:2021sqa}. The full amplitude is an interference of the longitudinal (``$0$'') and two transverse (``$+$'' and ``$-$'') terms. The couplings $H_{0,-,+}$ of the three amplitudes and the phases $\phi_{+,-}$ of the transverse couplings with respect to the longitudinal one have been measured to be
\begin{equation}
    \begin{split}
        |H_0| = & \phantom{-} 0.760 \pm 0.007 \pm 0.007, \\
        |H_-| = & \phantom{-} 0.195 \pm 0.022 \pm 0.032, \\
        |H_+| = & \phantom{-} 0.620 \pm 0.011 \pm 0.013, \\
        \phi_- = & -0.046 \pm 0.102 \pm 0.020\,\rad, \\
        \phi_+ = & \phantom{-} 0.108 \pm 0.170 \pm 0.051\,\rad. \\
    \end{split}
\end{equation}
The degree of parity violation is determined by phase differences $\phi_{\pm}$ and is consistent with zero in the measurements. However, in the simulation study, the maximal parity violation was introduced by taking $\phi_{\pm} = \pm\pi/2$. For the generated events, the same simulation and reconstruction procedure is applied as for the signal events, and the binned asymmetries then are calculated and fitted with the RH and PT templates. The results are presented in Fig.~\ref{fig:dstdsst_bck}, and the fitted couplings are given in Table~\ref{tab:bck_bias}. This background is mostly present in the low-$\cos\theta_D$ and low-$\cos\theta_{\ell}$ region, and even for the maximal parity violation the measured binned asymmetries are consistent with zero. This is not a surprise since the parity-odd term is effectively cancelled out after the integration over degrees of freedom related to the unreconstructed photon. 

The remaining double charm backgrounds are contributing at the level of 1\% or less. The most significant of them is likely to be $\B\to \Dstarp\D^{(*)-}_s\pi$ (where $\pi$ is either $\pi^0$ or $\pim$). It should mostly be mediated by the same $\D^{**}$ states as in the $\B\to D^{**}\mun\neumb$ decays. The evidence of this decay was reported by CLEO collaboration~\cite{CLEO:2000svj} and the branching ratio $\BR(\Bp\to \Db^{**0})\D_s^{(*)+}$ was measured to be $(2.73\pm 0.78\pm 0.48\pm 0.68)\%$. 
Under the assumptions that $\BR(\D^{**0}\to \Dstarp\pim)=1/3$ and $\BR(\Bz\to\D^{**-}\D_s^{(*)+}) = \BR(\Bp\to\D^{**0}\D_s^{(*)+})$, this background should contribute at around 2\% level. Similar decay modes with the $\D^{(*)}K$ combination, where either higher $\Ds$ excitations via external $W$ emission, or charmonium resonances via internal $W$ emission are produced, is another group of double charm backgrounds. Several such modes are observed~\cite{BaBar:2010tqo}, with each individual mode expected to contribute at the $(0.1$--$1)$\% level. 

The $\B\to \Dstarp\D^{(*)-}_s\pi$ and $\B\to \Dstarp\D^{(*)}K$ decays can exhibit the mechanism of parity and $CP$ violation similar to that discussed in Section~\ref{sec:sl_bck}, via the interference of multiple broad overlapping states. Unlike $\B\to \D^{**}\mun\neumb$ decays, they are potentially $CP$-violating in the SM. However, since these decays are dominated by the $\bquark\to \cquark$ transition, their $CP$-violating terms are suppressed at least as $|V_{ub}/V_{cb}|$. As a result, they are not expected to contribute to the $CP$ asymmetry in the $\Bzb\to \Dstarp\mun\neumb$ decays at the noticeable level. Their contributions, however, can be non-negligible for the $\Bzb\to \Dstarp\taum\neutb$ decays. More precise determination of the branching fractions of these doubly charmed modes and studies of their amplitude structure at $B$ factories and LHCb should help better constrain their effect. 

\subsection{Instrumentation effects}

In this section, the reconstruction effects that can introduce parity-odd terms to the distribution that is originally parity-even will be considered. The parity-odd terms are basically ``twists'' in the kinematic distribution of the decay that are asymmetric with respect to mirror reflection. Detector effects such as misalignments or non-uniform reconstruction efficiency that are mirror-asymmetric (chiral) may thus introduce such terms in the observable distribution. In real analysis, one has to develop a data-driven procedure to control these effects precisely. 

\subsubsection{Detector misalignment}

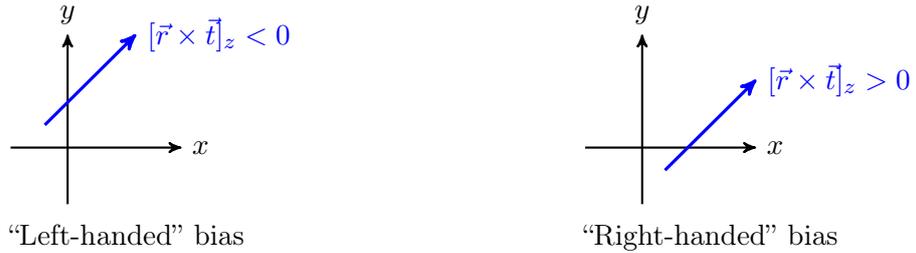
\begin{figure}
    \hfill
    \parbox[t]{0.3\textwidth}{
    \begin{tikzpicture}[scale=1.5, 
    axis/.style={thick, ->, >=stealth'}, 
    track/.style={very thick, ->, >=stealth'}
    ]
    \draw[axis] (-0.5,0) -- (1,0) node(xline)[right] {$x$};
    \draw[axis] (0,-0.5) -- (0,1) node(yline)[above] {$y$};
    \draw[track,blue] (-0.2, 0.2) -- (0.6, 1.0) node()[right] {$[\vec{r}\times \vec{t}]_z<0$}; 
    \end{tikzpicture}
    }
    \hfill
    \parbox[t]{0.3\textwidth}{
    \begin{tikzpicture}[scale=1.5, 
    axis/.style={thick, ->, >=stealth'}, 
    track/.style={very thick, ->, >=stealth'}
    ]
    \draw[axis] (-0.5,0) -- (1,0) node(xline)[right] {$x$};
    \draw[axis] (0,-0.5) -- (0,1) node(yline)[above] {$y$};
    \draw[track,blue] (0.2, -0.2) -- (1.0, 0.6) node()[right] {$[\vec{r}\times \vec{t}]_z>0$}; 
    \end{tikzpicture}
    }
    \hfill
    
    \hfill
    \parbox[t]{0.3\textwidth}{
    ``Left-handed'' bias
    }
    \hfill
    \parbox[t]{0.3\textwidth}{
    ``Right-handed'' bias
    }
    \hfill
  \caption{Illustration of ``left-handed'' and ``right-handed'' bias in 
  the track parameters relative to the original production point at $x=y=0$. }
  \label{fig:track_asym}
\end{figure}

Detector misalignment results in the systematic bias of the reconstructed parameters of the final state particles (in the case of the $\Bzb\to \Dstarp\mun\neumb$ analysis, of the four charged tracks) from the true ones. The effect of misalignment is expected to be more pronounced in the angular analysis of semileptonic decays compared to similar analyses with fully reconstructed decays (such as measurements of triple product asymmetries in four-body $\B$ hadron decays~\cite{LHCb:2018fpt, LHCb:2014djq}) since the former relies on reconstruction of vertices to obtain angular observables. Relatively small mirror-asymmetric misalignment of the elements of the tracking detectors can introduce biases in the parameters of a single track such that it is systematically biased ``left'' or ``right'' relative to its production point (see Fig.~\ref{fig:track_asym}). 

In the case of LHCb, the largest contribution to the parity-odd misalignment effect is expected to come from the vertex detector (VELO~\cite{Alves:2008zz}) since it is the detector element that provides the information about the impact parameters (IP) of the tracks and, as a consequence, the positions of decay vertices. VELO consists of two moving halves (left and right with respect to the beams) which are opened and closed at each beam injection. As a result, calibration of the relative positions of the two halves is the largest source of systematic uncertainty in the vertex position. These positions are calibrated for each fill with the precision of a few micrometres in the $x-y$ plane (transverse to the beams) and up to a few tens of $\mum$ along the beams~\cite{Aaij:2014zzy}. 

Misalignment of each VELO half is parametrised by six components: translations $T_{x,y,z}$ along the three axes and rotations $R_{x,y,z}$ around them. The global shifts of VELO as a whole do not introduce bias in the measurement of $\B$ decay vertex relative to the PV, therefore, only the translations and rotations of the two VELO halves in the opposite directions are considered in the following study. For the simulated signal decays, the 
10\mum translation and 10\murad rotation of each VELO half in the opposite directions is simulated by modifying the track parameters separately for the tracks with the negative ($t_x<0$) and positive ($t_x>0$) slopes. $\B$ decay vertices are refitted and the biases in the decay parameters $q^2$, $\cos\theta_{D}$, $\cos\theta_{\ell}$ and $\chi$ are calculated. Binned asymmetry histograms are then obtained for each misalignment component and fitted with the RH and PT templates to obtain the biases in the NP couplings. 
These are listed in Table~\ref{tab:align_bias}. 

\begin{table}
  \caption{Fitted values of the NP couplings from the binned asymmetry fit for the translational and rotational components of VELO halves misalignment}
  \label{tab:align_bias}
  \begin{center}
  \begin{tabular}{lrr}
\toprule
{Translation/rotation} & {$\Delta\Im(g_R)$} & {$\Delta\Im(g_P g_T^*)$} \\
\midrule
$T_x$ & $-0.00000 \pm 0.00020$ & $\left(-2.7 \pm 2.0\right) \times 10^{-5}$ \\
$T_y$ & $-0.00320 \pm 0.00019$ & $-0.000240 \pm 0.000019$ \\
$T_z$ & $\left(0.6 \pm 2.8\right) \times 10^{-5}$ & $\left(1.6 \pm 2.5\right) \times 10^{-6}$ \\
$R_x$ & $\left(-8.4 \pm 0.7\right) \times 10^{-5}$ & $\left(-5.3 \pm 0.6\right) \times 10^{-6}$ \\
$R_y$ & $\left(-9 \pm 6\right) \times 10^{-6}$ & $\left(3 \pm 5\right) \times 10^{-7}$ \\
$R_z$ & $\left(-1.3 \pm 0.5\right) \times 10^{-6}$ & $\left(-4 \pm 4\right) \times 10^{-8}$ \\
\bottomrule
\end{tabular}

  \end{center}
\end{table}

\begin{figure}
    \centering

    \includegraphics[width=0.48\textwidth]{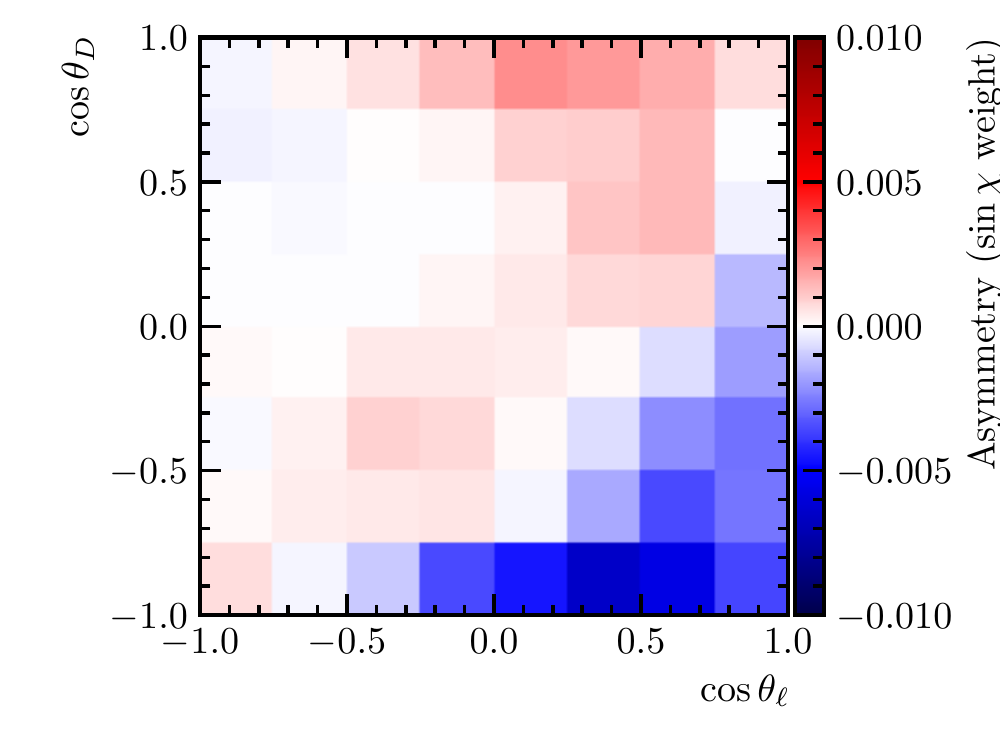}
    \put(-161, 131){\colorbox{white}{(a)}}
    \includegraphics[width=0.48\textwidth]{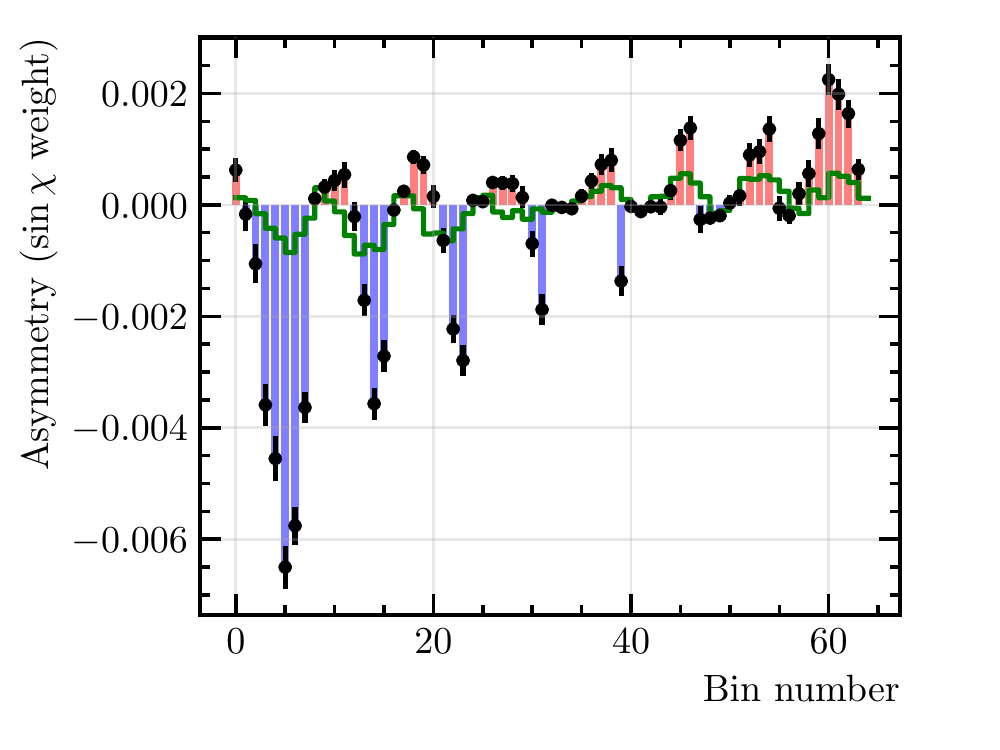}
    \put(-158, 131){(b)}

    \includegraphics[width=0.48\textwidth]{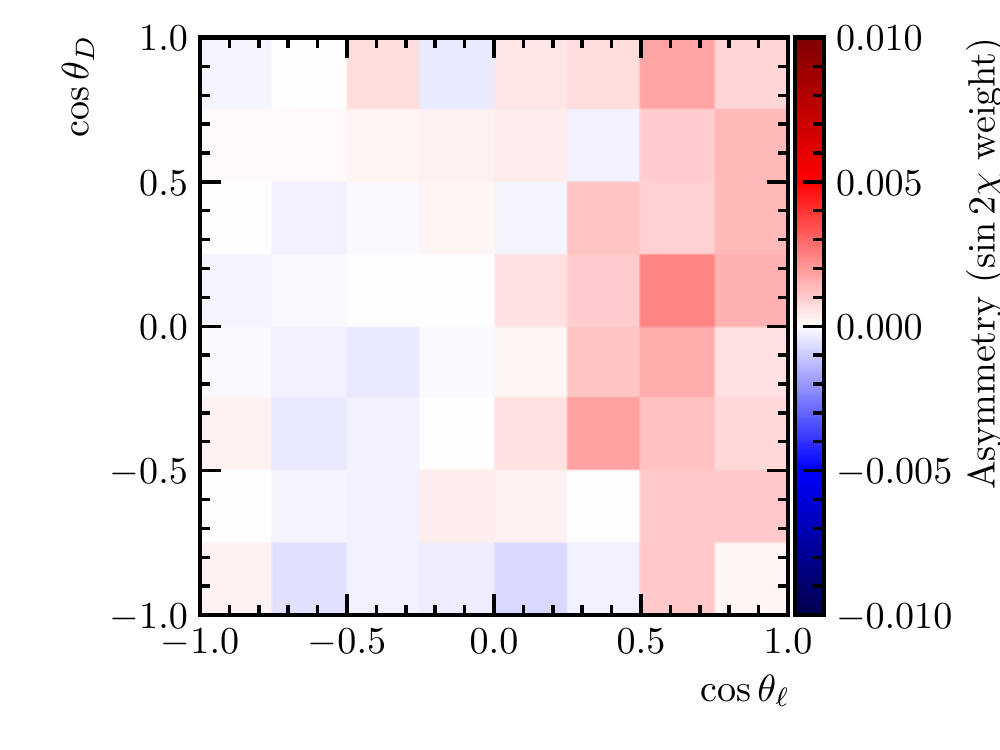}
    \put(-161, 131){\colorbox{white}{(c)}}
    \includegraphics[width=0.48\textwidth]{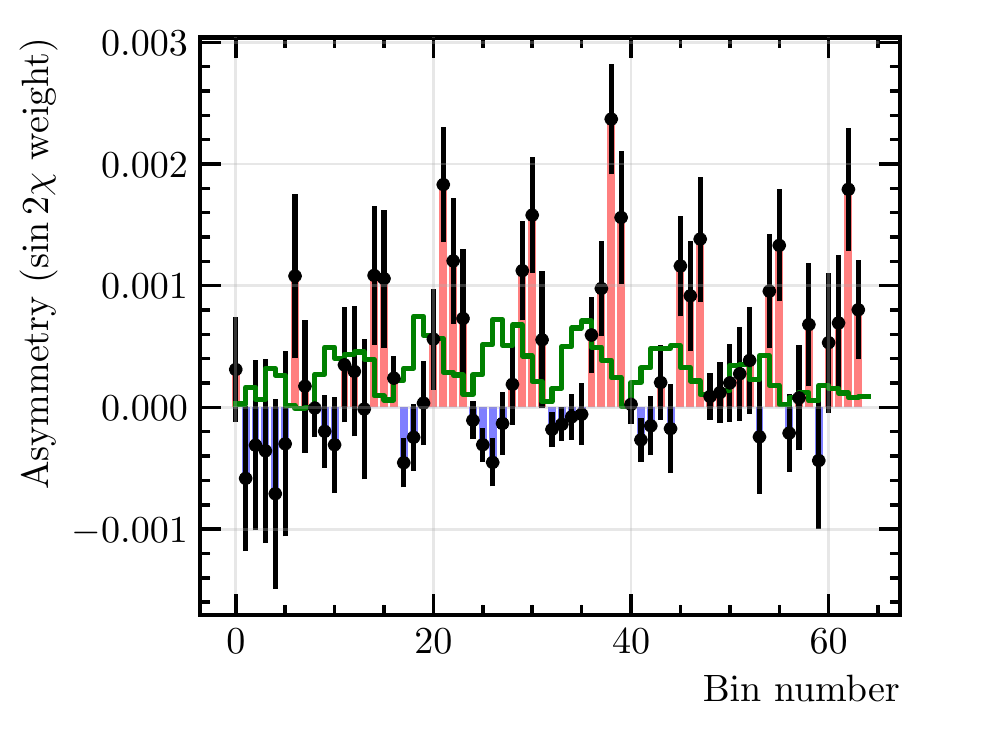}
    \put(-158, 131){(d)}

    \caption{Binned asymmetries of (a,b) $\sin\chi$ and (c,d) $\sin 2\chi$
    terms for the SM events with $T_y$ VELO misalignment of 10\mum. Plots (a) and (c) show 2D binned asymmetries, (b) and (d) are the corresponding flattened asymmetries (points with error bars), and the result of the best fit using RH and PT templates (solid-line histogram). }
    \label{fig:align_bias}
\end{figure}

As seen from Table~\ref{tab:align_bias}, the components that introduce significant nonzero $P$-odd terms are $T_y$ and $R_x$. In the case of $R_x$, the bias is numerically small, an order of magnitude smaller than the expected statistical precision in the 50\invfb sample. The bias introduced by the $T_y$ translation ($\pm 10$\mum misalignment for the left/right VELO halves), however, is of the order of the expected statistical precision. One can thus conclude that it is important to control the misalignment of VELO halves at the level of at least a few micrometres, which was shown to be achievable in the currently available data. Figure~\ref{fig:align_bias} shows the binned asymmetries due to $T_y$ misalignment and the fit result using RH and PT templates. The pattern of asymmetry as a function of $\cos\theta_D$ and $\cos\theta_{\ell}$ differs from that generated by NP terms, and thus can be independently controlled in the fit to data (\eg by introducing a dedicated component of asymmetry due to misalignment obtained with simulation). 

\subsubsection{Asymmetry in reconstruction efficiency}

Another possible instrumentation effect that could introduce chiral terms 
in the decay distribution is non-uniform reconstruction efficiency. The non-uniform efficiency in track reconstruction as a function of $\phi=\arctan(t_y/t_x)$ is not parity-odd since it averages out after integration over $\phi$. Still, there are at least three effects that are potentially parity-odd: 
\begin{itemize}
    \item {\bf Single-track efficiency terms}. 
    Non-zero parity-odd terms can appear in single-track efficiency if it depends on both the track direction $\vec{t}$ and coordinates of its origin $\vec{r}$. It should be proportional to the cross product $[\vec{r}\times \vec{t}]$. It is the only possible parity-odd effect if all tracks have the same efficiency dependence (\eg, for the reconstruction efficiency in VELO). 
In terms of $\sin\chi$ expansion~(\ref{eq:sinchi_xy}), this effect produces the terms proportional to $\sin(\phi_B - \phi_i)$ with $i$ being one of the charged final state particles. 
    \item {\bf Factorisable two-track efficiency terms}. For the track efficiency that is independent of origin position $\vec{r}$, parity-odd terms can still appear if the dependence on the direction $\vec{t}$ is different for different particle species. This effect is possible, \eg, in the contribution of particle identification subdetectors to the track reconstruction efficiency. In expansion~(\ref{eq:sinchi_xy}), this results in the terms proportional to $\sin(\phi_i-\phi_j)$, with both $i$ and $j$ being charged final states. 
    \item {\bf Non-factorisable two-track efficiency terms}. Finally, the parity-odd effects are possible in the non-factorisable contributions to the reconstruction efficiency, via the asymmetries in the opening angle $\Delta\phi$ between two tracks. They give rise to the same $\sin(\phi_i-\phi_j)$
terms in the $\sin\chi$ expansion (\ref{eq:sinchi_xy}) as the factorisable contributions above. 
\end{itemize}

Any of the parity-odd $\sin(\phi_i-\phi_j)$ terms can be multiplied by a parity-even function of the decay kinematics. Therefore, it is hard to define a general parametrisation of efficiency-related parity-odd effect, and it is not clear {\it a priori} which effects are dominant. In some cases, such as in the case of factorisable $\phi$-dependent efficiency, one can study the instrumentation effects using independent high-statistics calibration samples and then evaluate the effect for $CP$ violation measurement in $\Bzb\to \Dstarp\mun\neumb$ decays. However, one should keep in mind possible unaccounted effects, some of which could also be non-factorisable. Therefore, it is important to find a suitable control sample with properties similar to the signal decay that could be used to estimate the possible instrumentation effects. 

\subsubsection{Control sample}

The ideal control sample should satisfy the following conditions: 
\begin{itemize}
    \item be completely parity-even, without $P$-odd or $CP$-odd components even in reasonable NP scenarios; 
    \item have the same particles species in the final state as the signal mode;  
    \item have a significant overlap in the kinematic distributions and the same vertex topology; 
    \item have a larger yield than the signal mode in order for the measurement precision not to be dominated by systematic uncertainty. 
\end{itemize}
Probably the best approximation to the ideal control sample is the mode $\Bzb\to \Dp\mun\neumb$ with the subsequent $\Dp\to\Km\pip\pip$ decay. It has the same particles in the final state as the signal mode. Since it is a combination of two three-body decays, and the intermediate $D^+$ has zero spin, no $P$-odd effects are possible. Of the two pions in the final state, the one that gives the lower invariant mass combination $m(\Km\pip)$, can be used as a proxy for the ``slow'' pion from the $\Dstarp$ decay, the remaining $\Km\pip$ combination will then serve as a proxy for the $\Dz$ meson. Once the proxies for the $\Dstarp$ and ``slow'' pion are defined, the same phase space parameters $q^2$, $\cos\theta_D$, $\cos\theta_{\ell}$, and $\chi$ can be calculated. These will have little physical meaning, but the important point is that independent of the underlying dynamics of the $\Bz$ and $\Dp$ meson decays, the distribution of these variables should be purely $P$-even. 


$\Bzb\to \Dp\mun\neumb$ decays can be reconstructed by combining the  $\Dp\to\Km\pip\pip$ and muon candidates displaced from the primary vertex. In data, the pure signal will be contaminated by the decays with the excited charm states, \eg $\Bzb\to \Dstarp\mun\neumb$ with subsequent $\Dstarp\to \Dp\piz$ or $\Dstarp\to \Dp\gamma$ decays. However, even in the presence of NP that produces $CP$ violation in the $\Bzb\to \Dstarp\mun\neumb$ process, the decay parameters obtained from the reconstructed $\Dp\mun$ combinations are still $P$ and $CP$ conserving after the integration over the unreconstructed $\piz$ or $\gamma$, since the $\B$ and $\Dp$ decay densities factorise. This is checked using MC simulation. As a result, one can consider the decays $\Bzb\to \Dp\mun X$ as a control sample without the need to remove the candidates with $\Dp$ coming from excited charm states. This sample is dominated by $\Bzb\to \Dp\mun\neumb$ and $\Bzb\to \Dstarp\mun\neumb$, $\Dstarp\to \Dp\piz/\gamma$ processes, and the overall yield of such decays is expected to be about three times higher than that of the $\Bzb\to \Dstarp\mun\neumb$, $\Dstarp\to \Dz\pip$ decays assuming the same selection efficiency. 

\begin{figure}
  \centering
  \includegraphics[width=0.48\textwidth]{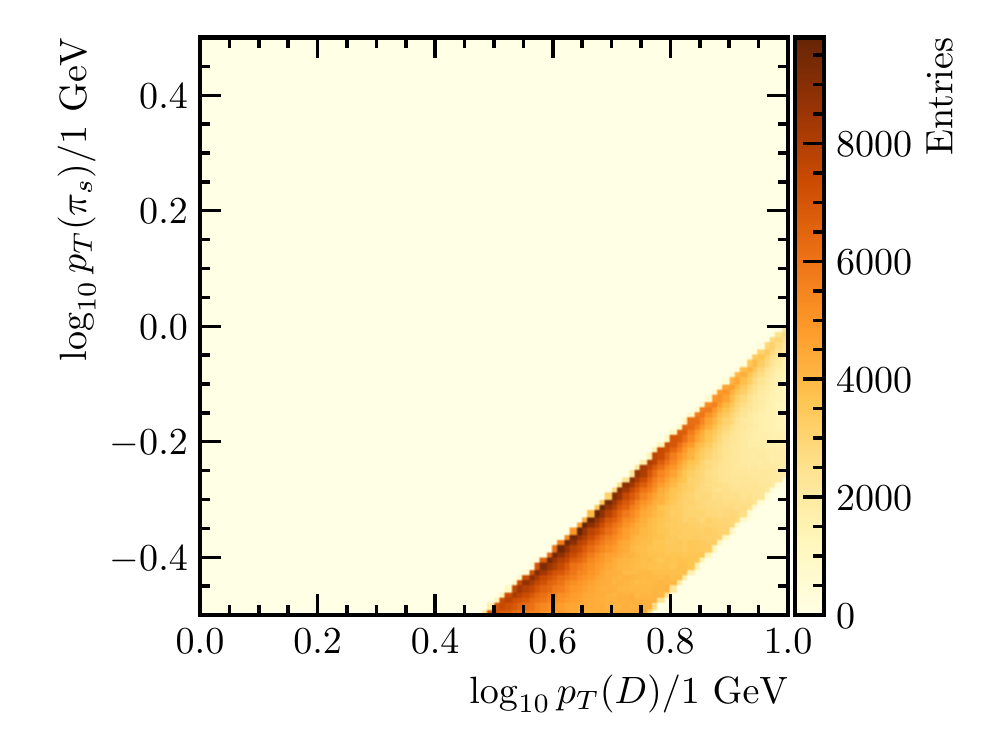}
  \put(-161, 131){\colorbox{white}{(a)}}
  \includegraphics[width=0.48\textwidth]{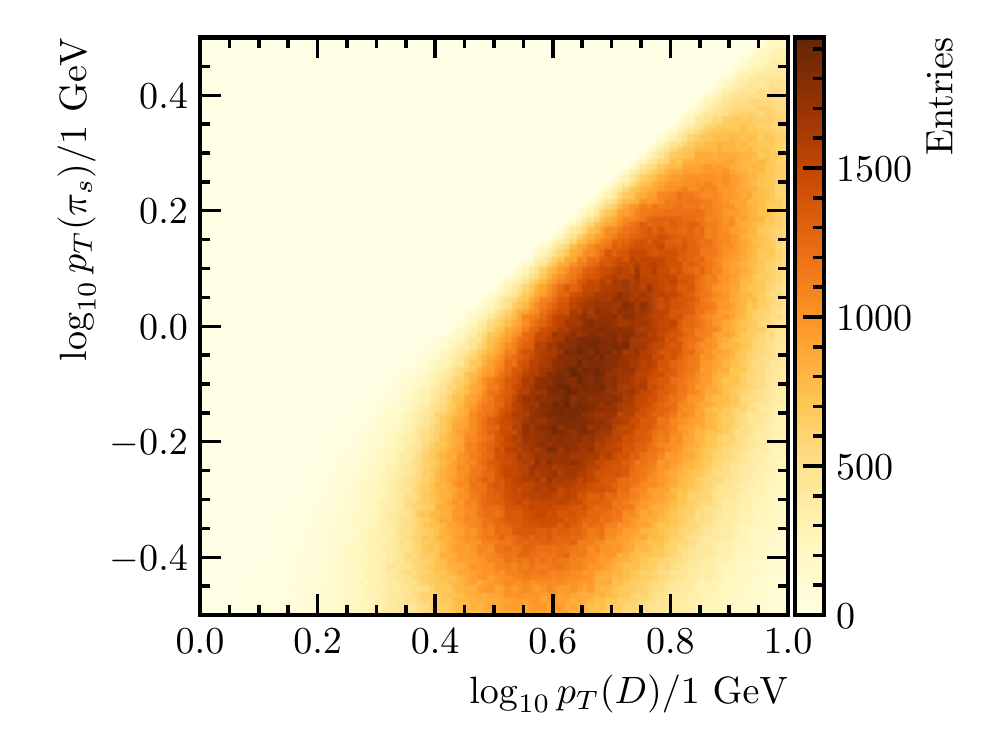}
  \put(-161, 131){\colorbox{white}{(b)}}
  \caption{Comparison of kinematic ranges in $\pt(\D)$ and $\pt(\pi_s)$ covered by the (a) signal sample $\Bzb\to \Dstarp\mun\neumb$ and (b) control sample $\Bzb\to \Dp\mun\neumb$. In the case of the control sample, the ``$D$'' is defined as the higher invariant-mass combination of the kaon and pion from $\Dp\to \Km\pip\pip$ decay, and ``$\pi_s$'' is the pion that forms the lower invariant-mass $\Km\pip$ combination.}
  \label{fig:dplnu_kine}
\end{figure}

The kinematic coverage of the control sample does not fully match that of the signal. The most notable difference is in the $\pt$ distributions of the $\Dz$
and $\pi_s$ (or their proxies in the case of the control sample; see Fig.~\ref{fig:dplnu_kine}). The $\pt(\pi_s)$ distribution tends to be harder for the control sample. The topology of the decay vertices is also different, with $\pi_s$ proxy being produced in the tertiary charm vertex rather than in the secondary $\B$ vertex as for the signal. Due to these differences, the possible systematic bias of NP couplings in the control channel cannot be taken directly as an estimate of the systematic uncertainty of the signal. Instead, the trends as a function of kinematic and topological variables have to be studied and extrapolated for the signal. 

\begin{figure}
  \vspace{-\baselineskip}
  \centering
  \includegraphics[width=0.48\textwidth]{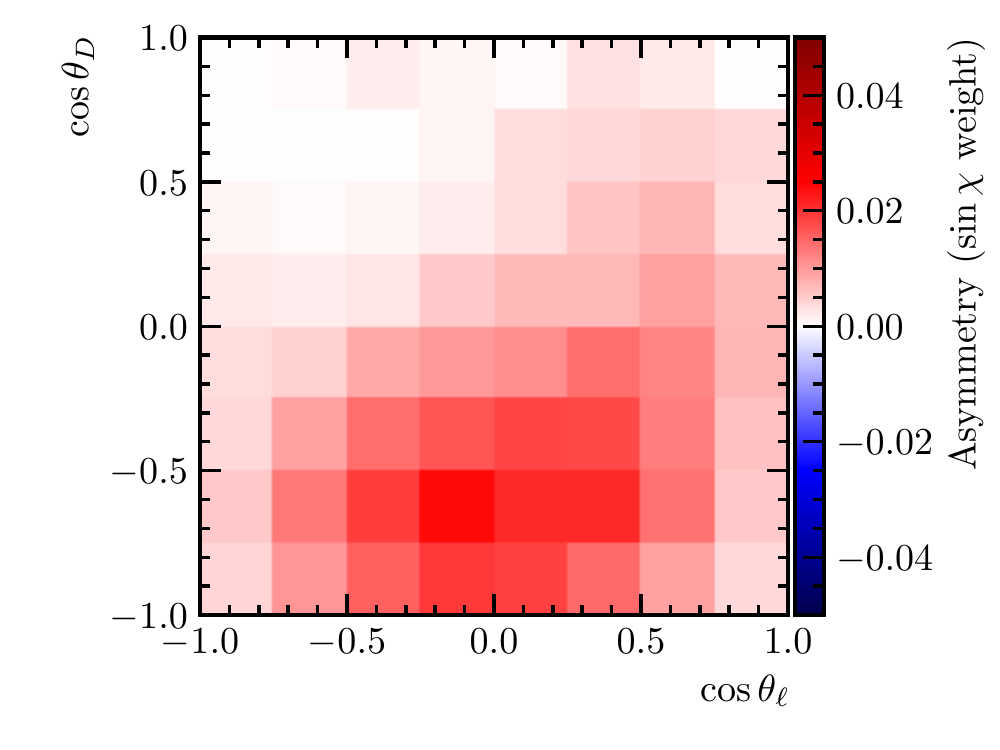}
  \put(-161, 131){\colorbox{white}{(a)}}
  \includegraphics[width=0.48\textwidth]{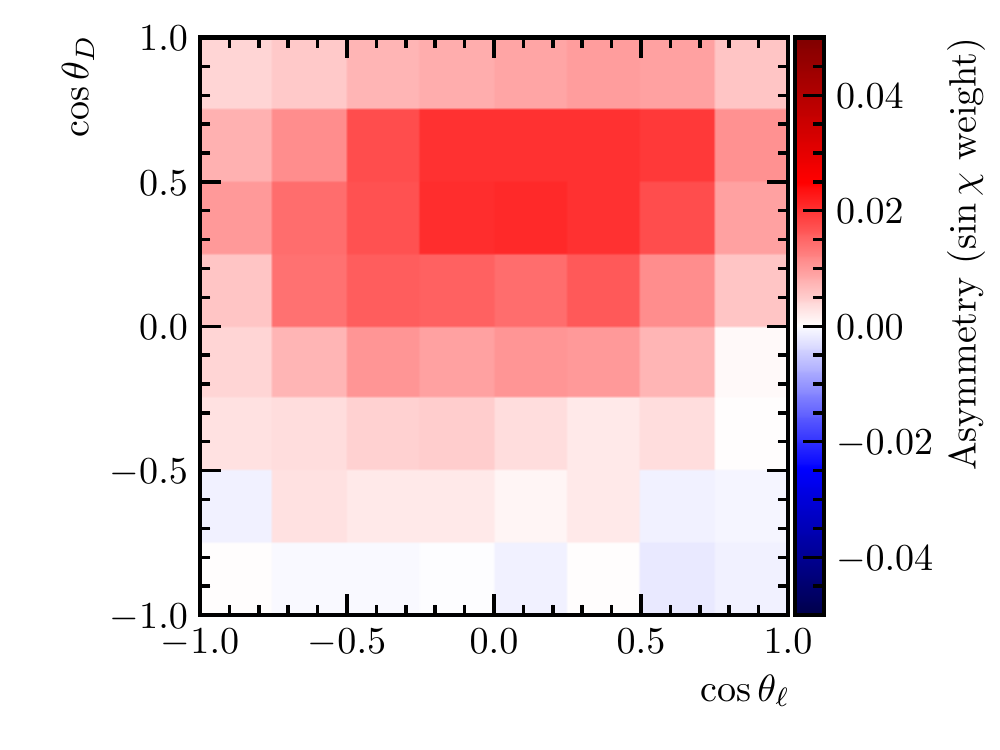}
  \put(-161, 131){\colorbox{white}{(b)}}

  \includegraphics[width=0.48\textwidth]{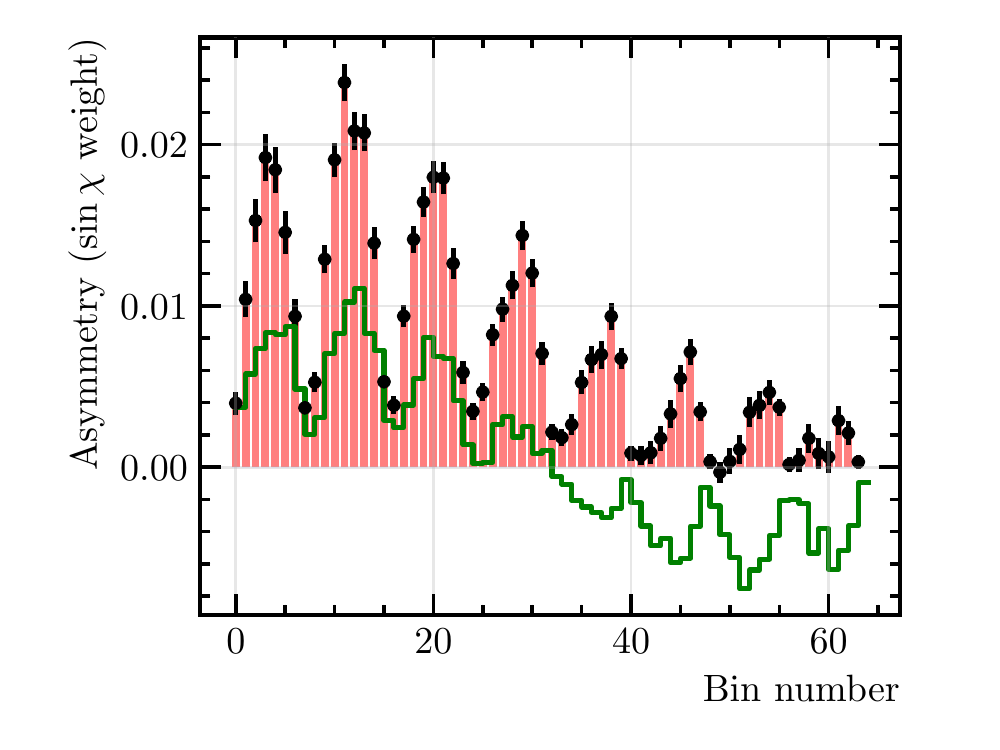}
  \put(-158, 131){(c)}
  \includegraphics[width=0.48\textwidth]{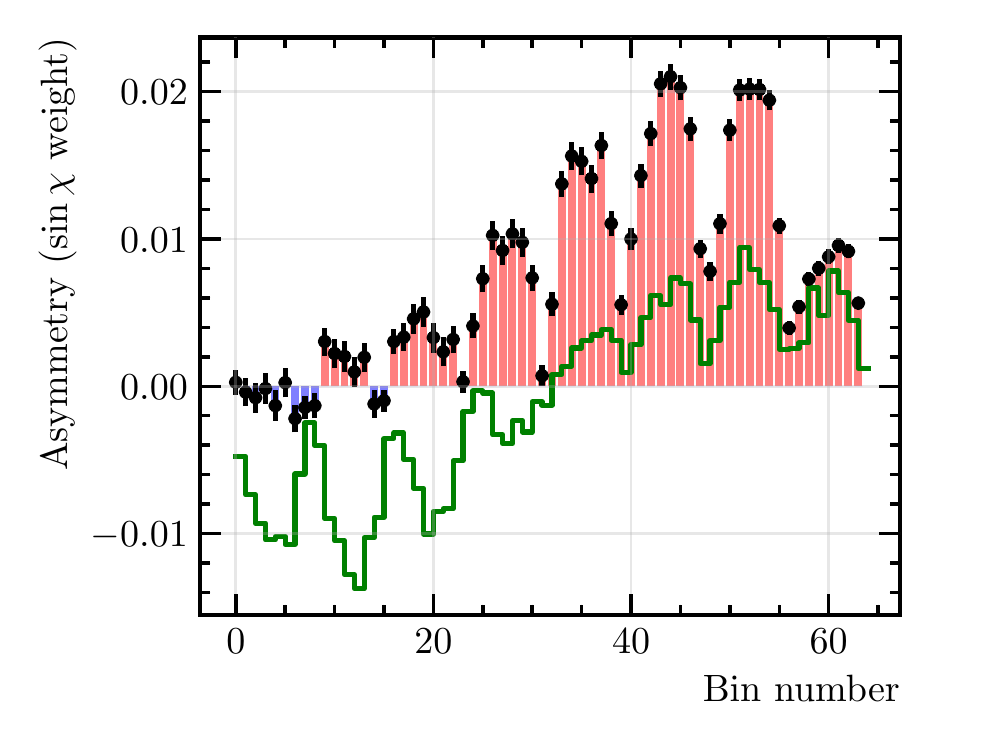}
  \put(-158, 131){(d)}

  \includegraphics[width=0.48\textwidth]{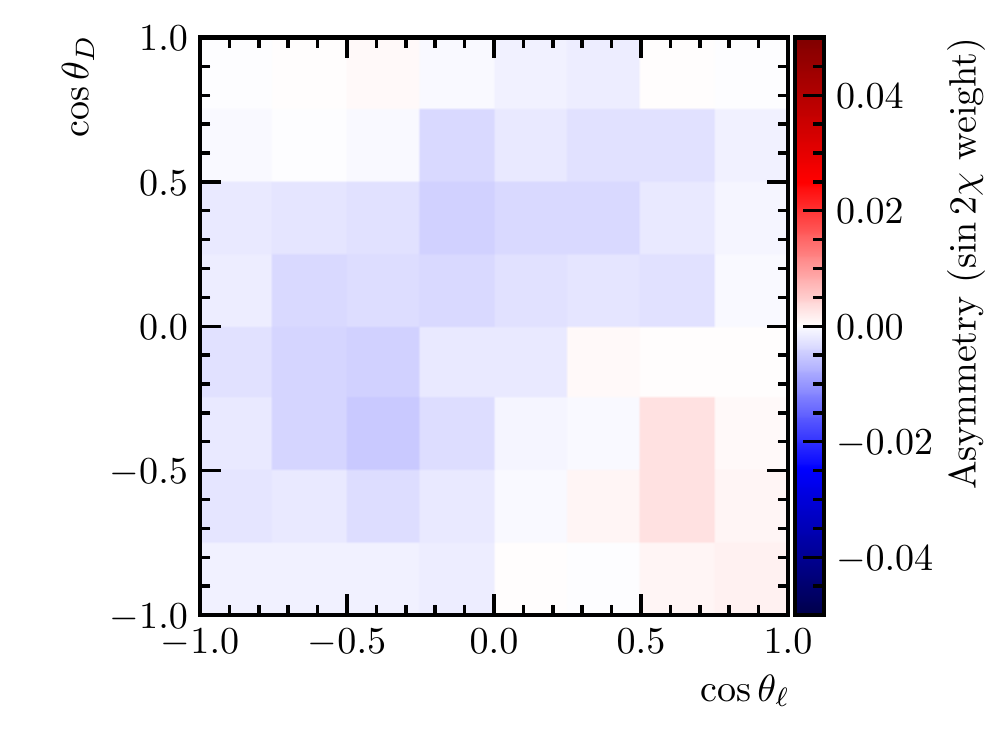}
  \put(-161, 131){\colorbox{white}{(e)}}
  \includegraphics[width=0.48\textwidth]{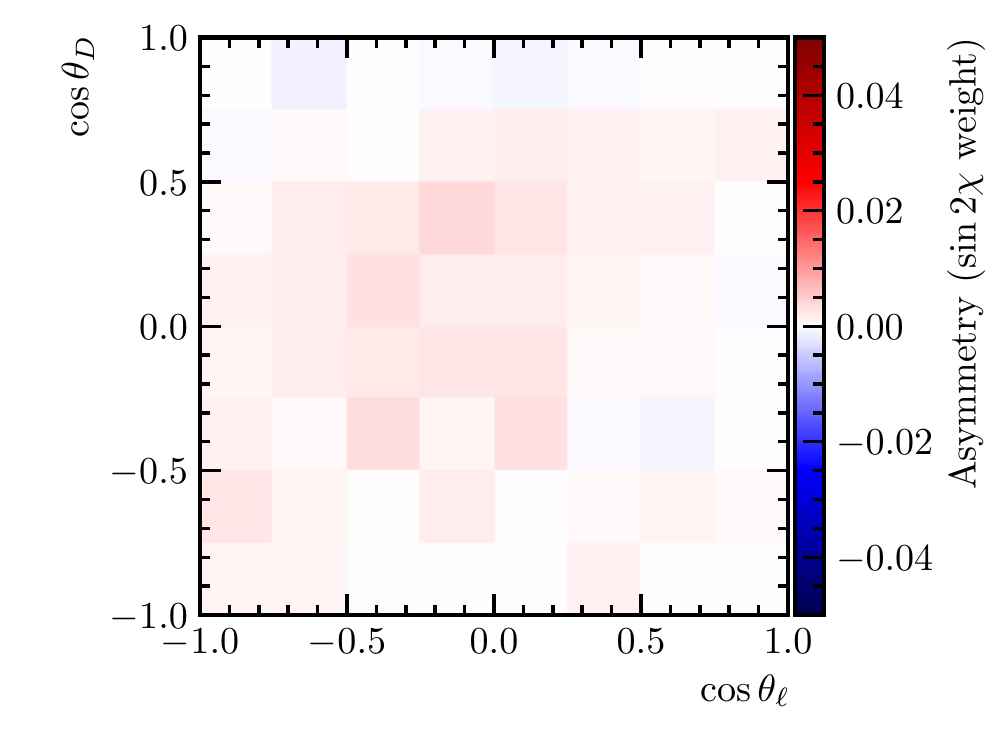}
  \put(-161, 131){\colorbox{white}{(f)}}
  
  \includegraphics[width=0.48\textwidth]{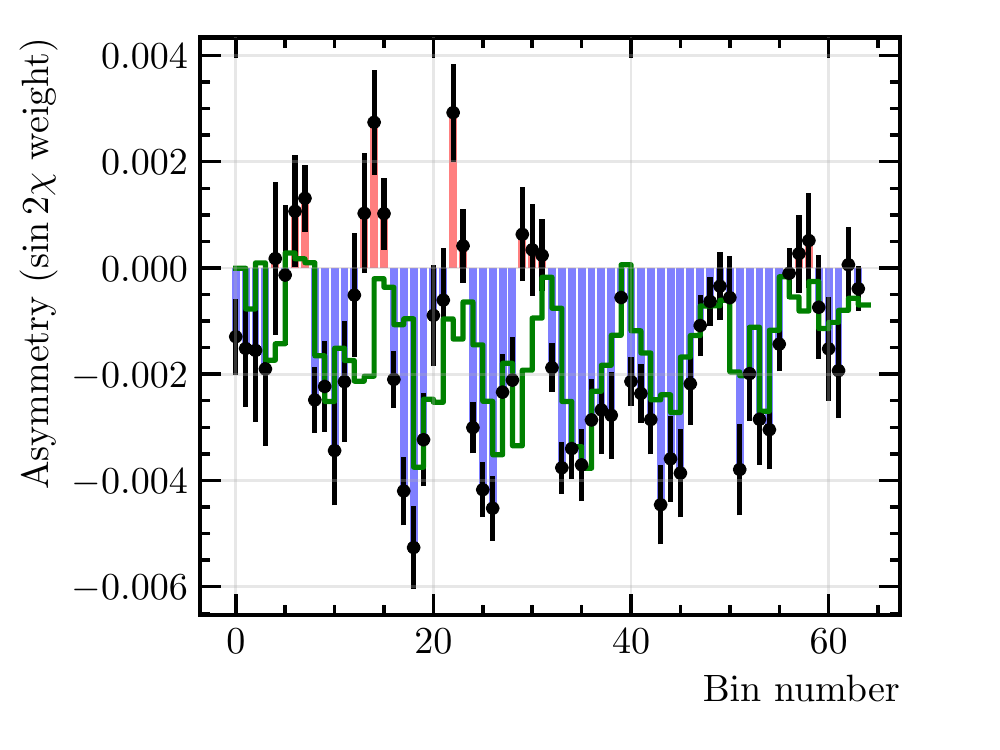}
  \put(-158, 131){(g)}
  \includegraphics[width=0.48\textwidth]{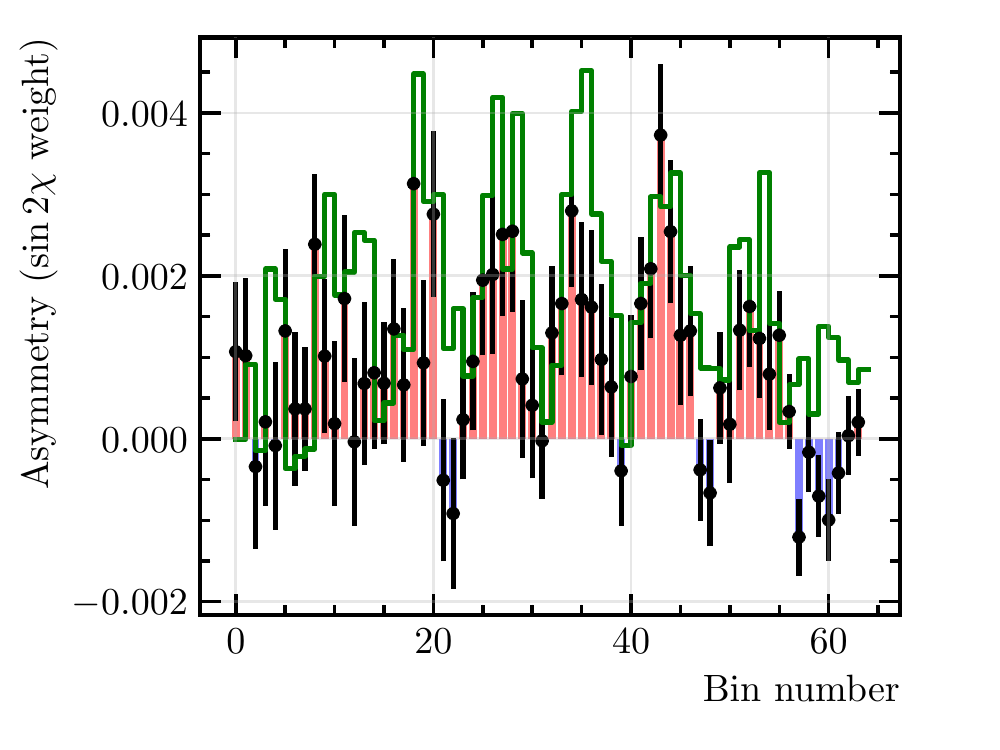}
  \put(-158, 131){(h)}

  \caption{Asymmetry due to parity-odd efficiency terms introduced by the $\phi$ dependence of reconstruction efficiency. Comparison of the (left plots) signal sample and (right plots) control sample. (a,b,e,f) 2D binned asymmetries and (c,d,g,h) the corresponding flattened asymmetries for (a--d) $\sin\chi$ and (e-h) $\sin2\chi$ asymmetry terms. The green solid line in plots (c,d,g,h) is the result of the template fit. }
  \label{fig:control}
\end{figure}

The effect of several parity-odd effects on the biases in the signal and control samples is demonstrated in Table~\ref{tab:control}. Non-uniform efficiency is simulated by applying a weight to each event. Since the effect of efficiency is linear and parity-even terms produce zero asymmetry, only the parity-odd weights are applied as listed below. The efficiency weights applied are $\mathcal{O}(1)$, which is much larger than the parity-odd efficiency variations that should appear in reality (if any). The five examples of systematic effects are: 
\begin{itemize}
  \item {\bf $T_y$ misalignment}: displacement of VELO halves ($+10\mum$ in $y$ direction for the left half and $-10\mum$ for the right half). 
  \item {\bf ``Left-right'' asymmetry in track efficiency}: 
  Efficiency weight $\varepsilon = {\rm sign}[\vec{r}\times \vec{t}]_z$. 
  \item {\bf $[\vec{r}\times \vec{t}]$ term in track efficiency}: 
  Efficiency weight $\varepsilon = [\vec{r}\times \vec{t}]_z/1\mm$. 
  \item {\bf $\phi$ dependence of track efficiency}: 
  Efficiency weight $\varepsilon = \sin\phi_{\mu} \cos\phi_{\pi_s}$. 
  \item {\bf $\Delta\phi$ dependence of track pair efficiency}: 
  Efficiency weight $\varepsilon = \sin(\phi_{\mu}-\phi_{\pi_s})$. 
\end{itemize}
One can see that the biases in the control samples are similar in magnitude to the ones of the signal. The biases due to efficiency terms, however, are opposite in sign. Figure~\ref{fig:control} shows the comparison of the asymmetries in bins of $\cos\theta_D$ and $\cos\theta_{\ell}$ for the signal and control samples produced by the $\phi$ dependence of track efficiency (by applying the weight $\varepsilon = \sin\phi_{\mu} \cos\phi_{\pi_s}$). The $\cos\theta_D$ dependence is different due to differences in kinematics of the two decay modes, and as a result, the sign of the $\sin 2\chi$ terms are opposite. What is important, however, is that the magnitudes of asymmetry (both the raw asymmetry in bins and the fitted values of the NP couplings) are similar, which permits the use of the control sample to evaluate the systematic uncertainty due to instrumentation effects in the signal fit. Because of the differences in some kinematic parameters, it might be instructive to measure the asymmetries in the control channel as a function of those parameters (\eg $\pt$ of the slow pion proxy) to better evaluate possible systematic biases for the signal kinematics. 

\begin{table}
  \caption{Comparison of NP coupling biases introduced by various instrumentation effects to the signal and control samples. See the text for definitions of the effects introduced. }
  \label{tab:control}
  \begin{center}
  \begin{tabular}{llrr}
\toprule
{Instrumentation effect} & {Coupling} & {Signal bias} & {Control bias} \\
\midrule
$T_y$ misalignment & $\Delta\Im(g_R)$ & $-0.00320 \pm 0.00019$ & $-0.00226 \pm 0.00018$ \\
 & $\Delta\Im(g_P g_T^*)$ & $-0.000240 \pm 0.000019$ & $-0.000378 \pm 0.000015$ \\
L-R asym. & $\Delta\Im(g_R)$ & $0.0377 \pm 0.0014$ & $-0.0312 \pm 0.0013$ \\
 & $\Delta\Im(g_P g_T^*)$ & $0.01050 \pm 0.00017$ & $-0.01414 \pm 0.00014$ \\
$\vec{r}\times \vec{t}$ term & $\Delta\Im(g_R)$ & $0.00155 \pm 0.00010$ & $-0.00149 \pm 0.00009$ \\
 & $\Delta\Im(g_P g_T^*)$ & $0.000381 \pm 0.000013$ & $-0.000554 \pm 0.000010$ \\
$\phi$ dependence & $\Delta\Im(g_R)$ & $0.0123 \pm 0.0006$ & $-0.0140 \pm 0.0006$ \\
 & $\Delta\Im(g_P g_T^*)$ & $0.00404 \pm 0.00008$ & $-0.00506 \pm 0.00006$ \\
$\Delta\phi$ dependence & $\Delta\Im(g_R)$ & $-0.0270 \pm 0.0007$ & $0.0281 \pm 0.0008$ \\
 & $\Delta\Im(g_P g_T^*)$ & $-0.00735 \pm 0.00010$ & $0.01077 \pm 0.00008$ \\
\bottomrule
\end{tabular}

  \end{center}
\end{table}

Other control samples can be used to address specific instrumentation effects. For instance, the VELO misalignment contribution, which is independent of the particle species, can be calibrated using clean high-statistics fully reconstructed samples such as $\Bp\to\jpsi\Kp$, $\jpsi\to\mup\mun$. The differences in the topology of vertices between the signal and control samples can be checked by comparing the $\Bzb\to\Dstarp\pim$ and $\Bzb\to\Dp\pim$ samples. In both cases, one can use a parity-odd observable such as the signed impact parameter of the $\B$ meson, $[\vec{r}_B\times \vec{t}_B]_z$, which should be zero on average but can deviate from zero due to experimental parity-odd effects. 


\section{Conclusion}
\label{sec:conclusion}

The possibility to perform the precision measurement of $CP$-violating observables in the $\Bzb\to \Dstarp\mun\neumb$ decays in proton-proton collisions at the LHCb detector is demonstrated. The feasibility study is performed using a simplified Monte Carlo simulation. The analysis can be performed in a way to cancel out the parity-even contributions to the decay density together with their corresponding uncertainties and concentrate only on the parity-odd (chiral) terms. As a result, theory uncertainties related to the formfactors and backgrounds can be kept low. 

The analysis allows to access two kinds of New Physics observables: imaginary part of right-handed coupling $\Im(g_R)$ and interference of pseudoscalar and tensor couplings $\Im(g_P g^*_T)$. Using the currently available LHCb data sample of 9\invfb one can constrain the contribution of right-handed current with the statistical uncertainty at the level of 1\% and pseudoscalar-tensor interference below 0.1\%. 

Systematic effects that could fake $CP$ asymmetry are considered. The contribution of backgrounds, which are mostly dominated by other semileptonic and double charm decays is evaluated to be small and well under control. The largest source of systematic bias is expected to be due to instrumentation effects such as misalignment of the detector elements or kinematic dependence of reconstruction efficiency. Parity-odd instrumentation effects can be controlled by the data samples which are inherently not $P$- or $CP$-violating, such as $\Bzb\to \Dp\mun\neumb$ with the subsequent $\Dp\to\Km\pip\pip$ decay. 

\acknowledgments

The authors are grateful to Aoife Bharucha and Jer\^{o}me Charles (CPT Marseille), and their colleagues from CPPM Marseille LHCb group and LHCb working group on semileptonic decays for useful discussions. 
This work received support from the French government under the France 2030 investment plan, as part of the Excellence Initiative of Aix-Marseille University - A*MIDEX (AMX-19-IET-008 - IPhU).


\bibliographystyle{JHEP}
\bibliography{paper}

\providecommand{\href}[2]{#2}\begingroup\raggedright\begin{thebibliography}{10}

\bibitem{BaBar:2012mrf}
{\bf BaBar} Collaboration, J.~P. Lees et~al., {\it {Measurement of branching
  fractions and rate asymmetries in the rare decays $B \to K^{(*)}\ell^+
  \ell^-$}},  {\em Phys. Rev. D} {\bf 86} (2012) 032012,
  [\href{http://arxiv.org/abs/1204.3933}{{\tt arXiv:1204.3933}}].

\bibitem{Belle:2016fev}
{\bf Belle} Collaboration, S.~Wehle et~al., {\it {Lepton-flavor-dependent
  angular analysis of $B\to K^\ast \ell^+\ell^-$}},  {\em Phys. Rev. Lett.}
  {\bf 118} (2017) 111801, [\href{http://arxiv.org/abs/1612.05014}{{\tt
  arXiv:1612.05014}}].

\bibitem{BELLE:2019xld}
{\bf Belle} Collaboration, S.~Choudhury et~al., {\it {Test of lepton flavor
  universality and search for lepton flavor violation in $B \rightarrow K\ell
  \ell$ decays}},  {\em JHEP} {\bf 03} (2021) 105,
  [\href{http://arxiv.org/abs/1908.01848}{{\tt arXiv:1908.01848}}].

\bibitem{Belle:2019oag}
{\bf Belle} Collaboration, A.~Abdesselam et~al., {\it {Test of lepton-flavor
  universality in ${B\to K^\ast\ell^+\ell^-}$ decays at Belle}},  {\em Phys.
  Rev. Lett.} {\bf 126} (2021) 161801,
  [\href{http://arxiv.org/abs/1904.02440}{{\tt arXiv:1904.02440}}].

\bibitem{LHCb:2014vgu}
{\bf LHCb} Collaboration, R.~Aaij et~al., {\it {Test of lepton universality
  using $B^{+}\rightarrow K^{+}\ell^{+}\ell^{-}$ decays}},  {\em Phys. Rev.
  Lett.} {\bf 113} (2014) 151601, [\href{http://arxiv.org/abs/1406.6482}{{\tt
  arXiv:1406.6482}}].

\bibitem{LHCb:2017avl}
{\bf LHCb} Collaboration, R.~Aaij et~al., {\it {Test of lepton universality
  with $B^{0} \rightarrow K^{*0}\ell^{+}\ell^{-}$ decays}},  {\em JHEP} {\bf
  08} (2017) 055, [\href{http://arxiv.org/abs/1705.05802}{{\tt
  arXiv:1705.05802}}].

\bibitem{LHCb:2019hip}
{\bf LHCb} Collaboration, R.~Aaij et~al., {\it {Search for lepton-universality
  violation in $B^+\to K^+\ell^+\ell^-$ decays}},  {\em Phys. Rev. Lett.} {\bf
  122} (2019) 191801, [\href{http://arxiv.org/abs/1903.09252}{{\tt
  arXiv:1903.09252}}].

\bibitem{LHCb:2019efc}
{\bf LHCb} Collaboration, R.~Aaij et~al., {\it {Test of lepton universality
  with $ {\Lambda}_b^0\to {pK}^{-}{\mathrm{\ell}}^{+}{\mathrm{\ell}}^{-} $
  decays}},  {\em JHEP} {\bf 05} (2020) 040,
  [\href{http://arxiv.org/abs/1912.08139}{{\tt arXiv:1912.08139}}].

\bibitem{LHCb:2021trn}
{\bf LHCb} Collaboration, R.~Aaij et~al., {\it {Test of lepton universality in
  beauty-quark decays}},  {\em Nature Phys.} {\bf 18} (2022) 277--282,
  [\href{http://arxiv.org/abs/2103.11769}{{\tt arXiv:2103.11769}}].

\bibitem{LHCb:2021lvy}
{\bf LHCb} Collaboration, R.~Aaij et~al., {\it {Tests of lepton universality
  using $B^0\to K^0_S \ell^+ \ell^-$ and $B^+\to K^{*+} \ell^+ \ell^-$
  decays}},  {\em Phys. Rev. Lett.} {\bf 128} (2022) 191802,
  [\href{http://arxiv.org/abs/2110.09501}{{\tt arXiv:2110.09501}}].

\bibitem{LHCb:2022qnv}
{\bf LHCb} Collaboration, R.~Aaij et~al., {\it {Test of lepton universality in
  $b \rightarrow s \ell^+ \ell^-$ decays}},
  \href{http://arxiv.org/abs/2212.09152}{{\tt arXiv:2212.09152}}.

\bibitem{BaBar:2012obs}
{\bf BaBar} Collaboration, J.~P. Lees et~al., {\it {Evidence for an excess of
  $\bar{B} \to D^{(*)} \tau^-\bar{\nu}_\tau$ decays}},  {\em Phys. Rev. Lett.}
  {\bf 109} (2012) 101802, [\href{http://arxiv.org/abs/1205.5442}{{\tt
  arXiv:1205.5442}}].

\bibitem{BaBar:2013mob}
{\bf BaBar} Collaboration, J.~P. Lees et~al., {\it {Measurement of an excess of
  $\bar{B} \to D^{(*)}\tau^- \bar{\nu}_\tau$ decays and implications for
  charged Higgs bosons}},  {\em Phys. Rev. D} {\bf 88} (2013) 072012,
  [\href{http://arxiv.org/abs/1303.0571}{{\tt arXiv:1303.0571}}].

\bibitem{Belle:2015qfa}
{\bf Belle} Collaboration, M.~Huschle et~al., {\it {Measurement of the
  branching ratio of $\bar{B} \to D^{(\ast)} \tau^- \bar{\nu}_\tau$ relative to
  $\bar{B} \to D^{(\ast)} \ell^- \bar{\nu}_\ell$ decays with hadronic tagging
  at Belle}},  {\em Phys. Rev. D} {\bf 92} (2015) 072014,
  [\href{http://arxiv.org/abs/1507.03233}{{\tt arXiv:1507.03233}}].

\bibitem{Belle:2016dyj}
{\bf Belle} Collaboration, S.~Hirose et~al., {\it {Measurement of the $\tau$
  lepton polarization and $R(D^*)$ in the decay $\bar{B} \to D^* \tau^-
  \bar{\nu}_\tau$}},  {\em Phys. Rev. Lett.} {\bf 118} (2017) 211801,
  [\href{http://arxiv.org/abs/1612.00529}{{\tt arXiv:1612.00529}}].

\bibitem{Belle:2017ilt}
{\bf Belle} Collaboration, S.~Hirose et~al., {\it {Measurement of the $\tau$
  lepton polarization and $R(D^*)$ in the decay $\bar{B} \rightarrow D^* \tau^-
  \bar{\nu}_\tau$ with one-prong hadronic $\tau$ decays at Belle}},  {\em Phys.
  Rev. D} {\bf 97} (2018) 012004, [\href{http://arxiv.org/abs/1709.00129}{{\tt
  arXiv:1709.00129}}].

\bibitem{Belle:2019rba}
{\bf Belle} Collaboration, G.~Caria et~al., {\it {Measurement of
  $\mathcal{R}(D)$ and $\mathcal{R}(D^*)$ with a semileptonic tagging method}},
   {\em Phys. Rev. Lett.} {\bf 124} (2020) 161803,
  [\href{http://arxiv.org/abs/1910.05864}{{\tt arXiv:1910.05864}}].

\bibitem{LHCb:2017smo}
{\bf LHCb} Collaboration, R.~Aaij et~al., {\it {Measurement of the ratio of the
  $B^0 \to D^{*-} \tau^+ \nu_{\tau}$ and $B^0 \to D^{*-} \mu^+ \nu_{\mu}$
  branching fractions using three-prong $\tau$-lepton decays}},  {\em Phys.
  Rev. Lett.} {\bf 120} (2018) 171802,
  [\href{http://arxiv.org/abs/1708.08856}{{\tt arXiv:1708.08856}}].

\bibitem{LHCb:2015gmp}
{\bf LHCb} Collaboration, R.~Aaij et~al., {\it {Measurement of the ratio of
  branching fractions $\mathcal{B}(\bar{B}^0 \to
  D^{*+}\tau^{-}\bar{\nu}_{\tau})/\mathcal{B}(\bar{B}^0 \to
  D^{*+}\mu^{-}\bar{\nu}_{\mu})$}},  {\em Phys. Rev. Lett.} {\bf 115} (2015)
  111803, [\href{http://arxiv.org/abs/1506.08614}{{\tt arXiv:1506.08614}}].
  [Erratum: Phys.Rev.Lett. 115, 159901 (2015)].

\bibitem{LHCb:2017rln}
{\bf LHCb} Collaboration, R.~Aaij et~al., {\it {Test of Lepton Flavor
  Universality by the measurement of the $B^0 \to D^{*-} \tau^+ \nu_{\tau}$
  branching fraction using three-prong $\tau$ decays}},  {\em Phys. Rev. D}
  {\bf 97} (2018) 072013, [\href{http://arxiv.org/abs/1711.02505}{{\tt
  arXiv:1711.02505}}].

\bibitem{Belle:2019ewo}
{\bf Belle} Collaboration, A.~Abdesselam et~al., {\it {Measurement of the
  $D^{\ast-}$ polarization in the decay $B^0 \to D^{\ast -}\tau^+\nu_{\tau}$}},
   in {\em {10th International Workshop on the CKM Unitarity Triangle}}, 3,
  2019.
\newblock \href{http://arxiv.org/abs/1903.03102}{{\tt arXiv:1903.03102}}.

\bibitem{Belle:2023bwv}
{\bf Belle} Collaboration, M.~T. Prim et~al., {\it {Measurement of Differential
  Distributions of $B \to D^* \ell \bar \nu_\ell$ and Implications on
  $|V_{cb}|$}},  \href{http://arxiv.org/abs/2301.07529}{{\tt
  arXiv:2301.07529}}.

\bibitem{Bobeth:2021lya}
C.~Bobeth, M.~Bordone, N.~Gubernari, M.~Jung, and D.~van Dyk, {\it
  {Lepton-flavour non-universality of ${\bar{B}}\rightarrow D^*\ell {{\bar{\nu
  }}}$ angular distributions in and beyond the Standard Model}},  {\em Eur.
  Phys. J. C} {\bf 81} (2021) 984, [\href{http://arxiv.org/abs/2104.02094}{{\tt
  arXiv:2104.02094}}].

\bibitem{Bhattacharya:2022bdk}
B.~Bhattacharya, T.~E. Browder, Q.~Campagna, A.~Datta, S.~Dubey, L.~Mukherjee,
  and A.~Sibidanov, {\it {Implications for the $\Delta A_{FB}$ anomaly in
  ${\bar B}^0\to D^{*+}\ell^- {\bar\nu}$ using a new Monte Carlo Event
  Generator}},  \href{http://arxiv.org/abs/2206.11283}{{\tt arXiv:2206.11283}}.

\bibitem{Becirevic:2019tpx}
D.~Be\v{c}irevi\'c, M.~Fedele, I.~Ni\v{s}and\v{z}i\'c, and A.~Tayduganov, {\it
  {Lepton Flavor Universality tests through angular observables of
  $\overline{B}\to D^{(\ast)}\ell\overline{\nu}$ decay modes}},
  \href{http://arxiv.org/abs/1907.02257}{{\tt arXiv:1907.02257}}.

\bibitem{Duraisamy:2013pia}
M.~Duraisamy and A.~Datta, {\it {The full $B \to D^{*} \tau^{-} \bar{\nu_\tau}$
  angular distribution and CP violating triple products}},  {\em JHEP} {\bf 09}
  (2013) 059, [\href{http://arxiv.org/abs/1302.7031}{{\tt arXiv:1302.7031}}].

\bibitem{Huang:2021fuc}
Z.-R. Huang, E.~Kou, C.-D. L\"u, and R.-Y. Tang, {\it {Un-binned angular
  analysis of $B\to D^*\ell \nu_\ell$ and the right-handed current}},  {\em
  Phys. Rev. D} {\bf 105} (2022) 013010,
  [\href{http://arxiv.org/abs/2106.13855}{{\tt arXiv:2106.13855}}].

\bibitem{Hill:2019zja}
D.~Hill, M.~John, W.~Ke, and A.~Poluektov, {\it {Model-independent method for
  measuring the angular coefficients of $B^0 \to D^{*-} \tau^+ \nu_{\tau}$
  decays}},  {\em JHEP} {\bf 11} (2019) 133,
  [\href{http://arxiv.org/abs/1908.04643}{{\tt arXiv:1908.04643}}].

\bibitem{Bhattacharya:2019olg}
B.~Bhattacharya, A.~Datta, S.~Kamali, and D.~London, {\it {CP violation in
  ${\bar B}^0\to D^{*+}\mu^-{\bar\nu}_\mu$}},  {\em JHEP} {\bf 05} (2019) 191,
  [\href{http://arxiv.org/abs/1903.02567}{{\tt arXiv:1903.02567}}].

\bibitem{Aloni:2018ipm}
D.~Aloni, Y.~Grossman, and A.~Soffer, {\it {Measuring CP violation in $b\to
  c\tau^-\bar{\nu}_\tau$ using excited charm mesons}},  {\em Phys. Rev. D} {\bf
  98} (2018) 035022, [\href{http://arxiv.org/abs/1806.04146}{{\tt
  arXiv:1806.04146}}].

\bibitem{Alves:2008zz}
{\bf LHCb} Collaboration, A.~A. Alves~Jr. et~al., {\it {The \lhcb detector at
  the LHC}},  {\em JINST} {\bf 3} (2008) S08005.

\bibitem{Ligeti:2016npd}
Z.~Ligeti, M.~Papucci, and D.~J. Robinson, {\it {New Physics in the visible
  final states of $B\to D^{(*)}\tau\nu$}},  {\em JHEP} {\bf 01} (2017) 083,
  [\href{http://arxiv.org/abs/1610.02045}{{\tt arXiv:1610.02045}}].

\bibitem{Cacciari:1998it}
M.~Cacciari, M.~Greco, and P.~Nason, {\it {The $p_T$ spectrum in heavy-flavor
  hadroproduction}},  {\em JHEP} {\bf 05} (1998) 007,
  [\href{http://arxiv.org/abs/hep-ph/9803400}{{\tt hep-ph/9803400}}].

\bibitem{Cowan:2016tnm}
G.~A. Cowan, D.~C. Craik, and M.~D. Needham, {\it {RapidSim: an application for
  the fast simulation of heavy-quark hadron decays}},  {\em Comput. Phys.
  Commun.} {\bf 214} (2017) 239--246,
  [\href{http://arxiv.org/abs/1612.07489}{{\tt arXiv:1612.07489}}].

\bibitem{Aaij:2014zzy}
R.~Aaij et~al., {\it {Performance of the LHCb Vertex Locator}},  {\em JINST}
  {\bf 9} (2014) P09007, [\href{http://arxiv.org/abs/1405.7808}{{\tt
  arXiv:1405.7808}}].

\bibitem{Dambach:2006ha}
S.~Dambach, U.~Langenegger, and A.~Starodumov, {\it {Neutrino reconstruction
  with topological information}},  {\em Nucl. Instrum. Meth. A} {\bf 569}
  (2006) 824--828, [\href{http://arxiv.org/abs/hep-ph/0607294}{{\tt
  hep-ph/0607294}}].

\bibitem{Gronau:2011cf}
M.~Gronau and J.~L. Rosner, {\it {Triple product asymmetries in $K$, $D_{(s)}$
  and $B_{(s)}$ decays}},  {\em Phys. Rev. D} {\bf 84} (2011) 096013,
  [\href{http://arxiv.org/abs/1107.1232}{{\tt arXiv:1107.1232}}].

\bibitem{PDG}
{\bf {Particle Data Group}} Collaboration, R.~L. Workman et~al., {\it {Review
  of Particle Physics}},  {\em Progress of Theoretical and Experimental
  Physics} {\bf 2022} (08, 2022)
  [\href{http://arxiv.org/abs/https://academic.oup.com/ptep/article-pdf/2022/8/083C01/45434166/ptac097.pdf}{{\tt
  https://academic.oup.com/ptep/article-pdf/2022/8/083C01/45434166/ptac097.pdf}}].
  083C01.

\bibitem{Bernlochner:2017jxt}
F.~U. Bernlochner, Z.~Ligeti, and D.~J. Robinson, {\it {Model independent
  analysis of semileptonic $B$ decays to $D^{**}$ for arbitrary new physics}},
  {\em Phys. Rev. D} {\bf 97} (2018) 075011,
  [\href{http://arxiv.org/abs/1711.03110}{{\tt arXiv:1711.03110}}].

\bibitem{BaBar:2010tqo}
{\bf BaBar} Collaboration, P.~del Amo~Sanchez et~al., {\it {Measurement of the
  $B\to \overline{D}{}^{(*)} D^{(*)}K$ branching fractions}},  {\em Phys. Rev.
  D} {\bf 83} (2011) 032004, [\href{http://arxiv.org/abs/1011.3929}{{\tt
  arXiv:1011.3929}}].

\bibitem{LHCb:2021sqa}
{\bf LHCb} Collaboration, R.~Aaij et~al., {\it {Angular analysis of $ {B}^0\to
  {D}^{\ast -}{D}_s^{\ast +} $ with $ {D}_s^{\ast +}\to {D}_s^{+}\gamma $
  decays}},  {\em JHEP} {\bf 06} (2021) 177,
  [\href{http://arxiv.org/abs/2105.02596}{{\tt arXiv:2105.02596}}].

\bibitem{CLEO:2000svj}
{\bf CLEO} Collaboration, S.~Ahmed et~al., {\it {Measurements of $B\to
  D_s^{(*)+} D^{*(*)}$ branching fractions}},  {\em Phys. Rev. D} {\bf 62}
  (2000) 112003, [\href{http://arxiv.org/abs/hep-ex/0008015}{{\tt
  hep-ex/0008015}}].

\bibitem{LHCb:2018fpt}
{\bf LHCb} Collaboration, R.~Aaij et~al., {\it {Search for CP violation using
  triple product asymmetries in $\Lambda^{0}_{b}\to pK^{-}\pi^{+}\pi^{-}$,
  $\Lambda^{0}_{b}\to pK^{-}K^{+}K^{-}$ and $\Xi^{0}_{b}\to pK^{-}K^{-}\pi^{+}$
  decays}},  {\em JHEP} {\bf 08} (2018) 039,
  [\href{http://arxiv.org/abs/1805.03941}{{\tt arXiv:1805.03941}}].

\bibitem{LHCb:2014djq}
{\bf LHCb} Collaboration, R.~Aaij et~al., {\it {Search for $CP$ violation using
  $T$-odd correlations in $D^0 \to K^+K^-\pi^+\pi^-$ decays}},  {\em JHEP} {\bf
  10} (2014) 005, [\href{http://arxiv.org/abs/1408.1299}{{\tt
  arXiv:1408.1299}}].

\end{thebibliography}\endgroup

\end{document}